\documentclass[pdflatex, sn-basic, 12pt]{sn-jnl} 

\usepackage{setspace}
\usepackage{graphicx}%
\usepackage{multirow}%
\usepackage{amsmath,amssymb,amsfonts}%
\usepackage{amsthm}%
\usepackage[title]{appendix}%
\usepackage{xcolor}%
\usepackage{textcomp}%
\usepackage{manyfoot}%
\usepackage{booktabs}%
\usepackage{algorithm}%
\usepackage{algorithmicx}%
\usepackage{algpseudocode}%
\usepackage{listings}%

\setcitestyle{aysep={ }} 

\begin{document}

\title[Article Title]{Prediction and Predictability of the Wet-Season Rainfall over Southeast India}

\author[1,2]{\fnm{Harini} \sur{S}} 

\author*[1,2]{\fnm{Devabrat} \sur{Sharma}} \email{devabratsharma9597@gmail.com}

\author[1,2]{\fnm{Yogenraj} \sur{Patil}}

\author[3]{\fnm{Gaurav} \sur{Chopra}} 

\author[1,2]{\fnm{Shruti} \sur{Tandon}}

\author[4]{\fnm{B. N.} \sur{Goswami}}

\author[1,2]{\fnm{R. I. } \sur{Sujith}} 

\affil[1]{\orgdiv{Department of Aerospace Engineering}, \orgname{Indian Institute of Technology Madras}, \orgaddress{\city{Chennai}, \state{Tamil Nadu}, \postcode{600036}, \country{India}}}

\affil[2]{\orgdiv{Centre for Excellence for studying Critical Transitions in Complex Systems}, \orgname{Indian Institute of Technology Madras}, \orgaddress{\city{Chennai}, \state{Tamil Nadu}, \postcode{600036}, \country{India}}}

\affil[3]{\orgdiv{Department of Applied Mechanics}, \orgname{Indian Institute of Technology Delhi}, \orgaddress{\city{Delhi}, \postcode{110016}, \country{India}}}

\affil[4]{\orgdiv{ST Radar Centre}, \orgname{Gauhati University}, \orgaddress{\city{Guwahati}, \state{Assam}, \postcode{781014}, \country{India}}}

\abstract{The challenge in predicting sub-regional climate within the Indian monsoon region is exacerbated by its increasing variability in a warming world. While exploring the seasonal predictability of rainfall over the state of Tamil Nadu in southeast India, we identify an overall increase in the monthly rainfall and its variability in recent years due to an increase in surface temperature, water vapour and moisture convergence. We attribute the increasing excess rainfall to a long-term reduction in convective inhibition. We further find an increasing trend in the length of the rainy season due to an earlier onset and a delayed withdrawal of the large-scale monsoon over the southeastern and southwestern regions of southern peninsular India, respectively. Further, the simultaneous (0-month lead) predictability of the primary wet-season (October-December, OND) rainfall over Tamil Nadu is dominated by sea surface temperature (SST) anomalies in the North Indian Ocean. However, a global tropical SST climate network reveals a high potential predictability and potential to realize significant forecast skill at a lead time of up to 10 months. The long-lead predictability arises from SST and rainfall interactions across the tropical Indo-Pacific and equatorial Atlantic regions. Our findings provide a robust data-driven methodology for skillful seasonal rainfall prediction over Tamil Nadu, despite the increasing rainfall variability.}



\maketitle
\section{Introduction}\label{Introduction}

Since more than 80\% of the annual rainfall over the Indian sub-continent occurs during June-September (JJAS), the all-India summer monsoon rainfall (ISMR) is often loosely referred to as the Indian monsoon rainfall. However, the remaining 20\% rainfall occurs during the Boreal winter season over two semi-arid regions of the country, namely, the south-east peninsula and north-west India \citep{parthasarathy1984interannual, dash2009changes, chakra2023multidecadal}. Also, a notable characteristic of the monsoon rainfall is that the interannual variability of seasonal mean rainfall is high where the mean is low and low when the mean is high \citep{parthasarathy1984interannual}. Therefore, the state of Tamil Nadu (approximately $8^{\circ}-13.5^{\circ}N$ and $76^{\circ}-80.5^{\circ}E$) located in southeast India, with low seasonal mean and high interannual variability, is highly vulnerable to climate change impacts \citep{varadan2017exploratory}. Hence, it is highly desirable to build a forecast model for predicting the peak wet-season (October-December, OND) rainfall over Tamil Nadu to build resilience against climate variability. In the present study, we propose the same while recognizing that low mean (signal) and high variability would make such a prediction challenging. As a prerequisite for building such a prediction model, we also examine the variability and predictability of OND rainfall over the south-east state of Tamil Nadu.

The seasonal cycle of Tamil Nadu rainfall (Fig. \ref{daily_clim_and_era_rf}a) is a small but integral part of the Indian monsoon rainfall. The Indian monsoon is characterized by the seasonal reversal of winds and the corresponding shift in rainfall distribution over the Indian landmass, reflecting the annual migration of the Intertropical Convergence Zone (ITCZ). The ITCZ is a zonally oriented rain band and a zone of intense convection and low-level convergence that shifts meridionally in response to large-scale atmospheric circulations. During Boreal summer (June to September or JJAS), the ITCZ migrates northward, creating strong cross-equatorial flow from the southern Indian Ocean and moisture-laden southwesterly winds from the Arabian Sea, resulting in the southwest monsoon over the Indian subcontinent. Conversely, during the Boreal winter (October to December or OND), the ITCZ retreats southward towards the equatorial Indian Ocean. This reversal of the circulation leads to northeasterly winds blowing towards the equatorial Indian Ocean associated with the northeast monsoon \citep{dhar1983foreshadowing, gadgil2003indian, loo2015effect, sreekala2018combined}. These winds pick up moisture primarily from the Bay of Bengal and deliver substantial rainfall to the southeastern coastal regions of India, particularly Tamil Nadu, coastal Andhra Pradesh, and parts of Kerala \citep{rajeevan2012northeast, sreekala2012northeast}. 

The southwest monsoon onset over Tamil Nadu occurs much later compared to most parts of India, typically after the monsoon has already advanced over the central and northern regions of the country \citep{gadgil2003indian, pai2020normal, patil2025climatic}. Owing to the barrier effect of the Western Ghats, most parts of Tamil Nadu are in the rain-shadow region during the southwest monsoon season, receiving only limited rainfall. Therefore, Tamil Nadu relies heavily on the northeast monsoon, which becomes the major contributor to its annual rainfall \citep{dhar1983foreshadowing, rajeevan2012northeast, prakash2013increasing, sreekala2018combined, lakshmi2021prediction}. The annual cycle of rainfall over Tamil Nadu is characterized by two seasonal (Fig. \ref{daily_clim_and_era_rf}a) peaks in rainfall and greater interannual variability \citep{rao1999variations, rajeevan2012northeast, sreekala2012northeast, krishnan2020assessment}. 

Variability of northeast monsoon rainfall in this region can also trigger droughts, severe water shortages, and crop failures, affecting millions in this region \citep{balachandran2006global, rajeevan2012northeast}. Further, the rapid urbanization and inadequate water storage infrastructure increase the impact of intense rainfall events, making the region highly sensitive to even small shifts in monsoon timing and intensity. In recent years, Tamil Nadu has witnessed an increase in daily extreme rainfall events and flooding during the northeast monsoon, such as the devastating Chennai floods during 2015, the widespread flooding during 2021, and more recent events associated with cyclonic activity, such as Cyclone Michaung in 2023 and Cyclone Fengal in 2024 \citep{prakash2013increasing, kartheeshwari20222021, sk2023decoding, shahi2023increase, aswini2024fengal, dev2025unravelling}. While the southwest monsoon of India has been widely studied \citep{krishnamurthy2000intraseasonal, goswami2006increasing, hrudya2021review}, only limited research focuses on the rainfall and the regions that receive major precipitation during the northeast monsoon season \citep{ zubair2006strengthening, rajeevan2012northeast, sreekala2012northeast, yadav2012enso}. 

Hence, understanding the dynamics of Tamil Nadu rainfall and its variability and predictability is crucial for building an appropriate seasonal rainfall prediction model optimizing agriculture production, and ensuring better resilience to climate variability under global climate change. Unlike the core monsoon zone of Central India, where the difference between peak and minimum rainfall is large ($\sim$10 mm/day, \cite{sharma2023variability}), Tamil Nadu shows a much smaller difference  ($\sim$3 mm/day; Fig. \ref{daily_clim_and_era_rf}a). This results in weaker seasonal contrast in Tamil Nadu, with the northeast monsoon dominating and the southwest monsoon also contributing to the annual rainfall.

Several studies have highlighted long-term trends and regional contrasts in rainfall over the Indian peninsular region influenced by both the southwest monsoon and the northeast monsoon \citep{rao1953study, kumar2004prediction, datta2022assessing, chakra2023multidecadal, rajasekaran2024quantifying}. The fluctuating relationship between the monsoon and other climatic phenomena such as El Ni\~no Southern Oscillation (ENSO) contributes to uneven rainfall distribution, and the recent trends point to a rise in extreme precipitation events, which are often linked to climate change \citep{rajeevan2012northeast, chen2019short, rajkumar2020extreme, sk2023decoding}. Theoretical and modelling studies indicate that atmospheric warming enhances moisture availability and the influence of physical drivers, which can increase extreme rainfall over the Indian region under favorable conditions \citep{goswami2006increasing, roxy2017threefold, ghausi2022breakdown, sengupta2023seasonal}. 

Despite these advances, there are very few studies that focus on the prediction and predictability of the northeast monsoon. Some studies have attempted the northeast monsoon prediction using statistical or machine-learning approaches based on sea surface temperature (SST) anomalies, general circulation models, and multi-model ensemble schemes \citep{acharya2011multi,rajeevan2012northeast, nair2013predictability, dash2019predictability}. \cite{lakshmi2021prediction} and \cite{tiwari2023evaluation} have examined the extremes, future projections and model evaluation of the northeast monsoon. However, most of these studies focus on either simultaneous or short lead forecasts, hindcast performance, or long-term projections. However, seasonal forecasts with little or no time to prepare for alternative agricultural planning are practically useless. Yet, the seasonal prediction of northeast monsoon rainfall, particularly at long lead times of several months, remains largely unexplored. In this study, we address the following questions: (i) What are the long-term changes in mean, variability, and extremes of rainfall over Tamil Nadu? (ii) What are the physical drivers of these changes? (iii) Are there changes in the monsoon season? and (iv) What is the predictability of OND rainfall over Tamil Nadu at long lead times?

To address these questions, we analyze the time series of rainfall over Tamil Nadu from 1940 to 2023 to understand its variability and potential changes in recent years due to climate change. We also examine the difference in monsoon onsets, withdrawals, and the length of the monsoon rainfall season (LRS) over southern India during the first and last 21 years for the time period 1940 to 2023. To determine the onset and withdrawal of the Indian monsoon, we utilize the large-scale onset approach developed by  \cite{patil2025climatic}, which views the arrival of monsoon as the emergence of large-scale organization from the clusters of local onsets, analyzed through complex network analysis. Since monsoon variability arises from complex multiscale interactions governing the climate system, the traditional identification of monsoon predictors based on direct correlation with regions of oceanic or atmospheric fields, although often showing significant relationships during the discovery phase, frequently performs poorly in independent forecast \citep{gadgil2005monsoon, wang2015rethinking}. 

The Earth's climate system is a complex system characterized by interactions across multiple spatial and temporal scales, comprising coupled ocean-atmosphere processes. Complex networks provide a useful framework to represent such systems, where nodes correspond to spatial locations, and links capture statistical or dynamical relationships between them. Climate networks have been increasingly used to study large-scale teleconnections and variabilities in the climate system \citep{donges2009complex, boers2019complex, fan2022network}. This motivates us to adapt and extend the complex network-based predictor discovery framework for seasonal monsoon rainfall developed by \cite{ran2025tropical}. Using the oceanic predictors for OND rainfall over Tamil Nadu identified through the complex network-based framework, we show that OND rainfall over Tamil Nadu has high predictability at a 10-month lead and demonstrate useful skill in independent out-of-sample seasonal forecasts.  

The rest of the paper is organized as follows. Section \ref{Data and Methods} describes the data and methodology, including data availability, climate network construction, predictor selection, and the ridge regression for the prediction framework. Section \ref{Results} presents the results, covering the variability and trends in rainfall over Tamil Nadu, atmospheric drivers of rainfall variability, changes in monsoon rainfall seasons, rainfall teleconnections with SST anomalies, and the predictability of OND rainfall of Tamil Nadu. Finally, Section \ref{Discussions} provides the discussions and conclusion.


\section{Data and Methods}\label{Data and Methods}
\subsection{Data}\label{Data}

For analyzing the monthly mean rainfall variabilities over Tamil Nadu, total precipitation (TP) data were obtained from multiple sources. These include hourly rainfall data from the fifth-generation reanalysis product ECMWF Reanalysis 5 (ERA5) of the European Centre for Medium-Range Weather Forecasts (ECMWF), gridded at $0.25^{\circ} \times 0.25^{\circ}$ \citep{hersbach2020era5} and available from 1940 to present; the India Meteorological Department (IMD) gridded daily rainfall data set developed by \cite{rajeevan2008analysis} at $1^{\circ} \times 1^{\circ}$ resolution and available from 1941 to present; and the high-resolution IMD hourly rainfall data set by \cite{pai2014development}, gridded at $0.25^{\circ} \times 0.25^{\circ}$ and available from 1901 to present. Hereafter, the IMD datasets of \cite{rajeevan2008analysis} and \cite{pai2014development} are referred to as IMD1 and IMD2, respectively. In this study, the temporal coverage of ERA5, IMD1, and IMD2 is 84 years, from 1 January 1940 to 31 December 2023. In addition, daily TP data were obtained from the Asian Precipitation–Highly Resolved Observational Data Integration Towards Evaluation of Water Resources (APHRODITE) data set, gridded at $0.25^{\circ} \times 0.25^{\circ}$, for the period 1951–2015 \citep{yatagai2012aphrodite}. 

To examine the dynamic and thermodynamic factors influencing the variability of monthly mean rainfall over Tamil Nadu, several atmospheric variables were obtained from ERA5 at $0.25^{\circ} \times 0.25^{\circ}$ resolution for the period 1940–2023. These include monthly surface temperature (ST), hourly total column water vapour (TCWV), hourly vertical integral of moisture convergence flux (VIMC), hourly convective available potential energy (CAPE), and hourly convective inhibition (CIN). The variable VIMC was calculated from the vertical integral of moisture flux divergence, and the 2m air temperature is represented as ST. For analyzing large-scale teleconnection patterns, monthly SST data over the global tropics ($0^{\circ}-360^{\circ}E$ and $30^{\circ}S-30^{\circ}NE$) gridded at $1^{\circ} \times 1^{\circ}$ from 1940-2023 are obtained from the Centennial in situ Observation-Based Estimate (COBE-SST2) dataset \citep{hirahara2014centennial}, while the hourly SST data for the same spatical coverage gridded at $0.25{^\circ}\times0.25{^\circ}$ for the same time period are obtained from ERA5. Hourly and daily products of TP, TCWV, VIMC, CAPE and CIN from ERA5, TP from IMD1, IMD2, and APHRODITE, and SST from COBE-SST2 were converted to a monthly time scale. Monthly anomalies were computed by removing the climatological mean of each calendar month from the corresponding monthly time series. All statistical analyses were performed using Python. 

\subsection{Methodology}\label{Methodology}
\subsubsection{Large-scale monsoon approach} \label{Large-scale monsoon approach}
To understand the changes in monsoon season over Tamil Nadu and surrounding regions, we analyze the large-scale onset and withdrawal dates that are derived through a novel complex network-based approach developed by \cite{patil2025climatic}. This approach views the arrival of the monsoon as the emergence of large-scale organization from the clusters of local onsets. A complex network consists of nodes that are connected using links. Here, the geographical locations are considered as nodes, and the links encode the information of connectivity of local onsets across the nodes. In this framework, on a given day, the links are established if two locations (nodes) have undergone a local onset and are spatial neighbours to each other. These links are removed when the locations undergo local withdrawal. The local onset and withdrawal are computed using the method of \cite{misra2018local}. 

In the local scale approach, to define onset and withdrawal, the local onset and withdrawal dates are identified using a daily accumulated rainfall anomaly series (from climatological annual mean) for each location across India. In this method, the minimum value of the anomaly curve represents the local onset date, and the maximum value provides the local withdrawal date. Using these local dates, a proximity network is constructed to identify clusters of local onsets. The first abrupt transition from the merging of two or more synoptic-scale clusters into a planetary-scale cluster is defined as the first large-scale onset over all locations of the cluster. Climatologically, the first large-scale onset occurs over Northeast India and progresses westward by adding new local onsets that join this cluster. Meanwhile, a synoptic scale cluster over peninsular India expands but remains separated from the westward expanding cluster of large-scale onsets. When these clusters merge, it results in a second abrupt transition, which defines the peninsular onset. The peninsular cluster advances northward and together with the westward expanding large-scale monsoon onset cluster, sets up a large-scale onset over the country. The first day on which a node detaches from the large-scale component is defined as the large-scale withdrawal date for that location. Finally, the difference between the large-scale onset and withdrawal dates provides LRS by the large-scale network approach.

\subsubsection{Climate Network Construction} \label{Network Construction}
Traditionally, the predictability of seasonal rainfall has been assessed using predictors identified from regions exhibiting the strongest correlation between the seasonal rainfall and large-scale fields such as SST. However, the key limitation of the traditional predictor discovery approach is that they are based solely on linear correlations. Such predictors may exhibit statistically significant relationships during the discovery phase; however, the prediction models built on them often fail during independent forecasts as they cannot adequately capture the complex interactions between the predictor and rainfall \citep{gadgil2005monsoon, wang2015rethinking}.

To overcome this hurdle, we construct complex networks using daily SST anomalies following a framework similar to that of \cite{ran2025tropical}, who developed a network based on daily 2m air temperature to generate predictors for forecasting summer rainfall over monsoon regions at lead times of 4–10 months. We use SST rather than surface air temperature as SST has a stronger control on the creation of the atmospheric heat source and hence the teleconnection.

To construct a complex network of daily SST anomalies over the global tropics, we obtained hourly SST from ERA5 reanalysis and created a daily mean. The data are provided on a global grid with a resolution of $0.25^{\circ} \times 0.25^{\circ}$ and are available from 1940 to the present. To reduce the computational cost during network construction, the daily SST fields are subsequently spatially averaged onto a coarser $2^{\circ} \times 2^{\circ}$ grid, by bilinear interpolation.

To efficiently utilize the 84 years of observational SST data for predictor discovery and model development, we split the dataset into two periods: 1940-1980 and 1981-2023. The split was designed to ensure that an independent verification of the seasonal monsoon forecasts, based on the network-based predictors, could be demonstrated over a sufficiently long recent period to establish robustness. Daily SST anomalies for the period 1940-1980 are computed relative to a day-wise climatology derived from the same period. Specifically, the anomaly for calendar day $d$ in year $y$ is defined as:
\begin{equation}
{SST}^{\prime}_{y}(d) =
{SST}_{y}(d) - \frac{1}{N_{1940-1980}(d)}
\sum_{i=1940}^{1980} {SST}_{i}(d) \label{eq1}
\end{equation}
Here, ${SST}_{y}(d)$ denotes the daily ${SST}_{y}(d)$ on the calendar day $d$ of the year $y$, and ${SST}^{\prime}_{y}(d)$ is the corresponding daily SST anomaly. The climatological mean is calculated separately for each calendar day $d$ by averaging over all available years in the period 1940-1980, with $N_{1940-1980}(d)$ representing the number of day $d$ within the period 1940-1980.

To ensure that the period 1981-2023 remains independent for assessing out-of-sample prediction skill of the model constructed using network-based predictors, and to avoid the use of future information in the anomaly calculation, daily SST anomalies beyond 1980 are computed using a progressively expanding climatological baseline. Accordingly, daily SST anomalies for a given year $y$ (where $y > 1980$) are calculated using data from 1940 up to the year $y$.
\begin{equation}
{SST}^{\prime}_{y}(d) =
{SST}_{y}(d) - \frac{1}{N_{1940-y}(d)}
\sum_{i=1940}^{y} {SST}_{i}(d) \label{eq2}
\end{equation}

For consistency in the daily climatology and to avoid unequal sampling, leap days (Feb 29) are excluded from the analysis, resulting in a fixed 365-day calendar for all years. Linear trends are removed from all anomaly data at each grid location.

Following the framework of \cite{ran2025tropical}, we construct a directed complex network using time-lagged Pearson correlations of daily SST anomalies \citep{donges2009complex, fan2021statistical, ran2025tropical}. Each $2^{\circ} \times 2^{\circ}$ grid box, across the global tropics is treated as a node in the network, resulting in a total of $N$ = 5400 nodes. (A detailed description of the network construction methodology is provided in Section 1.3 of the supplementary material).

For each year $y$ (1941 to 2023), correlations were computed between all node pairs for time lags $\tau = 0-200$ days, where the maximum lag of 200 ensures a reliable estimate of the background noise level. This produces 201 adjacency matrices of size $N \times N$ for each year $y$, where the elements of the adjacency matrix are defined as
\begin{equation}
{S^{y}_{i,j}(\tau) = Corr(SST^{y}_{i}(t-\tau), \,SST^{y}_{j}(t))} \label{eq4}
\end{equation}
Here, $i$ and $j$ represent two arbitrary nodes, $Corr(.)$ denotes the Pearson correlation coefficient, and $t$ spans the daily time indices from Jan 1 to Dec 31 (365 days). 

A matrix element ${S^{y}_{i,j(\tau)}}$ quantifies the extent to which SST variability at node $i$ precedes similar variability at node $j$ after a time delay $\tau$. Accordingly, outgoing links from a node indicate locations where SST anomalies tend to occur earlier and subsequently influence other regions, while incoming links identify locations whose variability tends to respond later to remote SST signals.

We identify SST regions that influence other regions by examining both the strength and time lag of their interactions. For each ordered pair of nodes $(i,j)$ we extract:
\begin{enumerate}[1.]
\item The maximum positive correlation across all lags - max(${S^{y}_{i,j}(\tau^{+})}$) and its corresponding time lag $\tau^{+}$.

\item The minimum negative correlation across all lags - min (${S^{y}_{i,j}(\tau^{-})}$)- and its corresponding time lag $\tau^{-}$.
\end{enumerate}

Only links with $\tau$ ($\tau^{+}$ and $\tau^{-}$) greater than zero are retained to avoid instantaneous correlations. Each non-zero element from this network is standardized to yield a weighted positive 2-D adjacency matrix ($WPA$) and a weighted negative 2-D adjacency matrix ($WNA$). Since the lag information is no longer retained, the direction of influence is determined by comparing the strength of interactions between each pair of nodes. For a given pair $(i,j)$, if the standardized positive link from $i$ to $j$ is stronger than that from $j$ to $i$, the interaction is considered to be directed from $i$ to $j$. Otherwise, the direction is reversed. This procedure defines the anti-symmetric matrix for the positive network. The same approach is used for the negative network. 

In network terminology, the number of links originating from a node is referred to as out-degree, while the number of links pointing toward a node is referred to as in-degree. Nodes with higher out-degree, therefore, act as source regions, exerting stronger dynamical influence on other nodes. As our goal is to identify such influential regions, we focus on outgoing connections. The out-degree of node $i$ is given by the total number of nodes $j$ for which the antisymmetric matrix of the positive or negative correlation network is equal to 1. Together with the corresponding link $WPA$ (or $WNA$), we compute the average strength of outgoing links for each node, separately for the positive complex network ${PCN^{out}_{i}(y)}$ and negative complex network ${NCN^{out}_{i}(y)}$. The node-wise metrics ${PCN^{out}_{i}(y)}$ and ${NCN^{out}_{i}(y)}$ each of length $N$ can be mapped back onto the original $2^{\circ} \times 2^{\circ}$ SST grid, yielding two spatial fields of size 30 x 180 for each year (Fig. S2). 

\cite{ran2025tropical} proposed a framework that selects only a single node from either the positive (${PCN^{out}}$) or negative (${NCN^{out}}$) complex climate network as a predictor of seasonal monsoon rainfall, based on the maximum correlation between the network nodes and the monsoon rainfall system. In contrast, our study utilizes all nodes from ${PCN^{out}}$ and ${NCN^{out}}$ that exhibit a statistically significant relationship with OND rainfall over Tamil Nadu across the entire tropical basin. Since OND rainfall over Tamil Nadu in any given season is influenced by the simultaneous contribution of multiple potential drivers distributed throughout the tropical basin, incorporating all statistically significant predictors (or nodes) from the complex network framework is essential for obtaining a more realistic estimate of predictability and for developing a skillful and robust independent forecast system.

Moreover, \cite{ran2025tropical} developed a linear regression model using network-derived predictors selected based on their correlation with seasonal monsoon rainfall over different regions. These models also demonstrated high forecast skill, with correlation coefficients ranging from 0.63 to 0.81 \citep{ran2025tropical}. However, one caveat is that the predictor selection for model development in \cite{ran2025tropical} relies on correlation information from both training and testing periods. As a result, reported skill reflects potential predictability rather than an independent out-of-sample forecast verification. In the present study, we explicitly demonstrate the feasibility of estimating both the long-lead potential predictability and the independent out-of-sample forecast skill for seasonal monsoon rainfall using network-derived predictors.

To integrate information from both positively and negatively influencing nodes, we further combine information from ${PCN^{out}_{i}(y)}$ and ${NCN^{out}_{i}(y)}$ by computing the long-term mean outgoing link strength at each grid box separately for the positive and negative correlation networks. For each year, a single complex network ${CN^{y}_{i}}$ is then constructed by selecting, at each grid box, the network (positive or negative) with the larger mean outgoing strengths.
\begin{equation}
CN^{y}_{i} =
\begin{cases}
PCN^{out}_{i},  & if \, \left| \overline{PCN^{out}_{i}} \right| > \left| \overline{NCN^{out}_{i}} \right|, \\[6pt]
NCN^{out}_{i}, & otherwise.
\end{cases}     \label{eq10}
\end{equation}
where $\left| \overline{PCN^{out}_{i}} \right| $ and $\left| \overline{NCN^{out}_{i}} \right| $ are the absolute values of long-term mean of $PCN^{out}_{i}(y)$ and $NCN^{out}_{i}(y)$ over the period 1941-2023, respectively.

\subsubsection{Predictor Selection} \label{Predictor Selection}

To demonstrate the applicability of complex network ($CN^{y}_{i}$) for long-lead seasonal monsoon forecasting, we focus on the northeast monsoon over Tamil Nadu, which is defined by the seasonal rainfall accumulated during the OND months. Since $CN^{y}_{i}$ is constructed using SST anomalies from year $y$, it can be used to forecast OND rainfall anomalies in the subsequent year $y+1$, corresponding to a lead time of approximately 10 months. Hence, the 10-month lead between the predictor and predictand arises by construction. As $CN^{y}_{i}$ is based on SST anomalies, the selected nodes represent ocean regions that can influence atmospheric conditions and moisture availability, and hence affect rainfall over Tamil Nadu. To develop a long-lead forecast model for OND rainfall anomalies, predictor selection is carried out in three steps:

(i) The $CN^{y}_{i}$ for the period 1941-2023 are correlated with OND rainfall anomalies for 1942-2024. Nodes that exhibit statistically significant correlation at the 95\% confidence level, based on p-values of the correlation, are retained and passed to the next step. Statistical significance at the 95\% confidence level implies that there is less than a 5\% probability that the observed correlation occurs due to random chance.

(ii) The selected nodes are further filtered using a phase-locking value (PLV; Fig. S3). PLV is a nonlinear measure to identify nodes that are in phase with each other, providing a more robust selection criterion associated with limited data \citep{lachaux1999measuring}. PLV is computed as
\begin{equation}
{PLV} = \left| \frac{1}{N} 
\sum_{t=1}^{N}
e^{i\,\Delta\phi(t)} \right|, \label{eq14}
\end{equation}
where $\Delta\phi(t) = \phi_1(t) - \phi_2(t)$, and $\phi_1(t)$ and $\phi_2(t)$ are the instantaneous phases of two signals.  

Nodes exhibiting statistically significant PLV values at the 95\% confidence level are retained for spatial clustering. If two or more such nodes are directly connected by neighbourhood adjacency, only the node with the maximum absolute correlation within the resulting cluster is retained and passed to the next step. This avoids redundancy arising from network clustering.

(iii)  To ensure predictor independence for the development of the forecast model, the selected nodes are correlated pairwise. If the correlation between two predictor nodes exceeds 0.7, one of the two predictor nodes is removed. Model complexity must be reduced when data availability is limited.

\begin{figure*}
    \centering
    \includegraphics[width=1\linewidth]{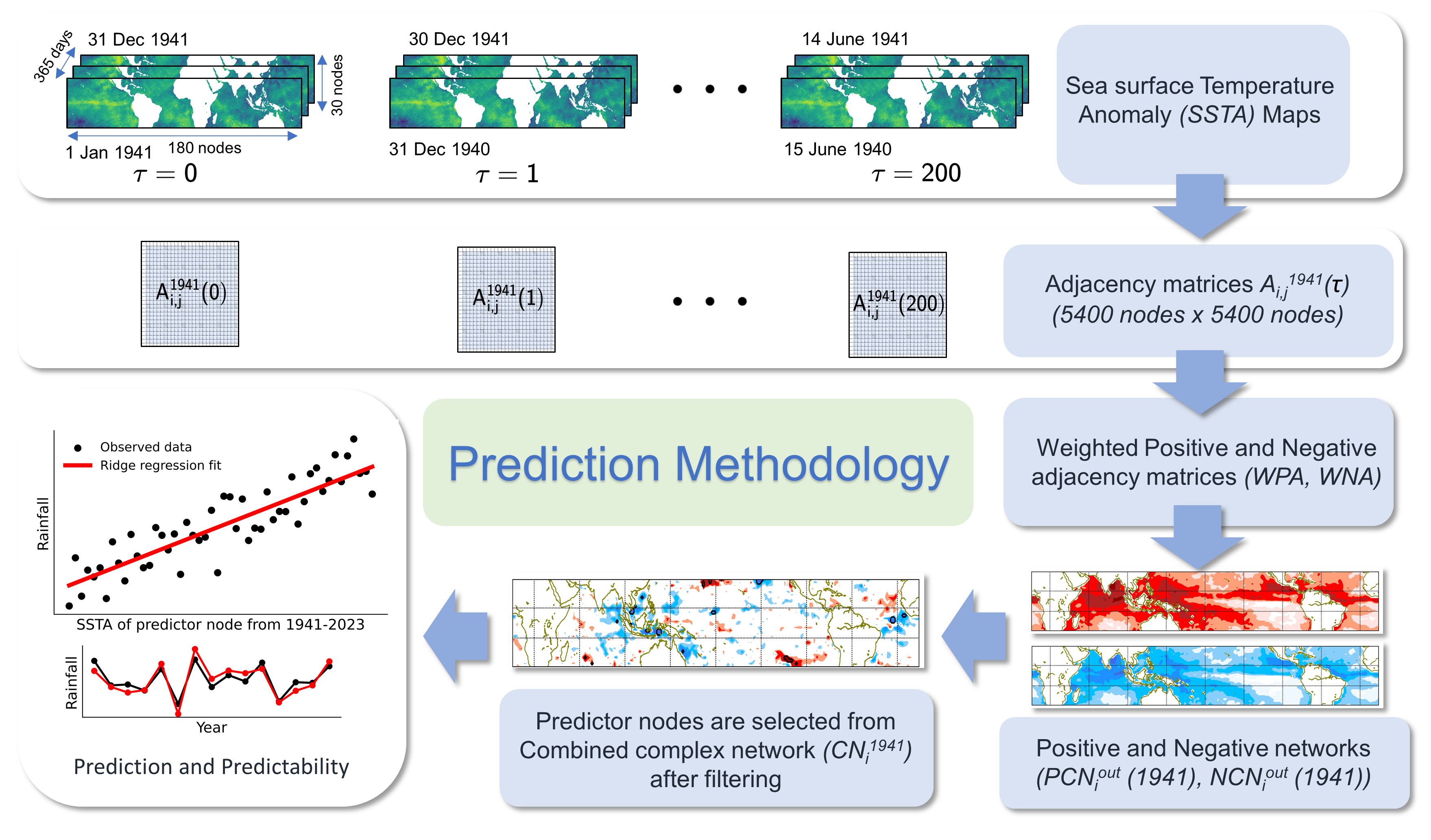}
    \caption{A schematic representation of network construction and forecast strategy}
    \label{schm_fig}
\end{figure*}

The selected predictor nodes through the three-step procedure are used to develop a ridge regression model for estimating the potential predictability and independent forecast skill of OND rainfall anomalies. This is schematically shown in Fig. \ref{schm_fig}.

\subsubsection{Ridge Regression} \label{Ridge Regression}

Due to the limited length of the observational dataset available for seasonal forecasting (84 years), developing machine learning models based on the predictors identified from a complex network framework is not feasible, as such models typically require large datasets for effective optimization. Moreover, deep learning without prior feature selection also requires large datasets to automatically identify relevant predictors before the final prediction stage. Hence, the complex network framework efficiently performs predictor selection using the available daily observations. Therefore, during independent testing, we employ ridge regression because of its simplicity and lower data requirement. In contrast to linear regression, ridge regression incorporates regularization to reduce over fitting and stabilize coefficient estimates, thereby limiting its dependence on the training data to some extent \citep{mcdonald2009ridge, hoerl2020ridge}. 

Let $y_t$ denote the OND rainfall anomaly in year $t$, and let $x_{t}= (x_{t,1},x_{t,2}, \ldots ,x_{t,p})$ represent the corresponding selected predictor nodes. The linear regression model is expressed as:
\begin{equation}
y_{t} = \beta_{\circ} + 
\sum^{p}_{j=1} \beta_{j} x_{t,j} +
\varepsilon_{t},
\end{equation}
where $\beta_{\circ}$ and $\beta_{j}$ are the intercept and regression coefficient, respectively, and $\varepsilon_{t}$ is the error term. The error can also be written as:
\begin{equation}
\varepsilon_{t} = y_{t} - \hat{y_{t}}, \label{eq16}
\end{equation}
where $\hat{y_{t}}$ is the predicted OND rainfall anomaly in year $t$. 

For $N$ years of data, the total loss function in linear regression is given by:
\begin{equation}
J(\beta) = 
\sum^{N}_{t=1} (y_{t} - \hat{y_{t}})^2 \label{eq17}
\end{equation}

In ridge regression, the coefficients are estimated by minimizing a penalized least-squared loss function:
\begin{equation}
J(\beta) = 
\sum^{N}_{t=1} (y_{t} - \hat{y_{t}})^2
+ \lambda \sum^{p}_{j=1} \beta^{2}_{j}, \label{eq18}
\end{equation}
where $\lambda = 10$ is a regularization parameter, chosen based on the maximum potential predictability estimate. The additional penalty term leads to better generalization of the model compared to linear regression.


\section{Results}\label{Results}
\subsection{Variability and trends in rainfall over Tamil Nadu} \label{Variability of monthly mean rainfall over Tamil Nadu}

To characterize the temporal evolution of rainfall over Tamil Nadu, we analyze the monthly rainfall data for the time period 1940-2023, from multiple sources, focusing on long-term changes in monthly rainfall variability. This approach allows us to examine the rainfall behavior of Tamil Nadu and provides a foundation for understanding the changes occurring in recent decades. 

\begin{figure*}
    \centering
    \includegraphics[width=1\linewidth]{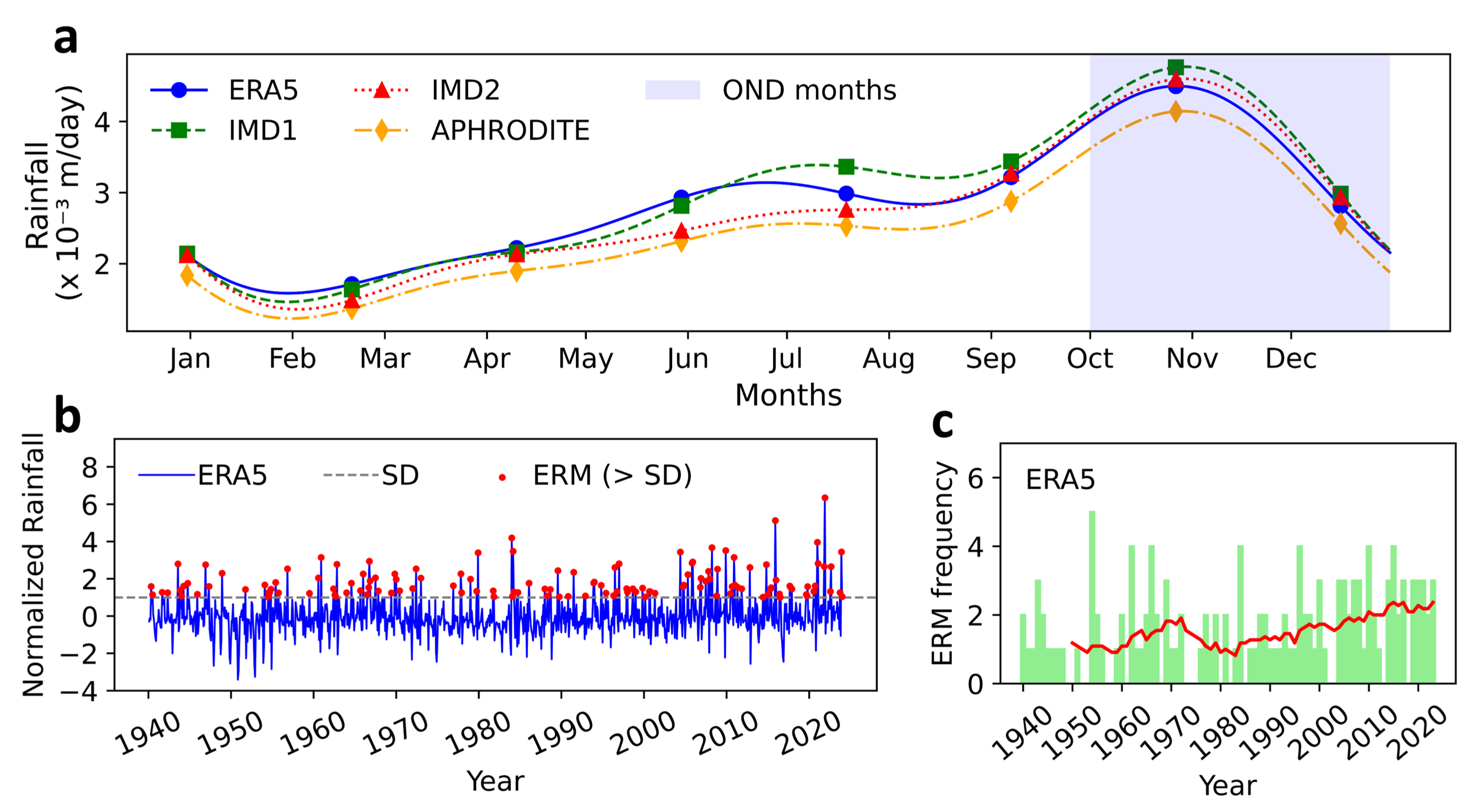}
    \caption{(a) Mean annual cycle of daily rainfall over Tamil Nadu derived from different rainfall datasets. (b) Monthly mean anomaly time series from ERA5, with red points indicating months with total monthly rainfall greater than +1 standard deviation (SD; marked in dashed gray line). (c) Count of extreme rainfall months (ERM) per year, derived from ERA5, is shown by counting the number of months in each year with rainfall greater than +1 SD. The red curve denotes the 11-year running mean. 
    }
    \label{daily_clim_and_era_rf}
\end{figure*}

The monthly anomaly rainfall time series using the ERA5 data set is presented in  Fig. \ref{daily_clim_and_era_rf}b. The normalized rainfall is obtained by dividing the monthly rainfall anomaly by the standard deviation (SD) of the monthly anomaly rainfall time series. High variability of monthly rainfall is apparent over Tamil Nadu. The months with rainfall larger than one SD are considered monthly extremes and are referred to as extreme rainfall months (ERM). ERM frequency is the count of such months in a year. The 11-year moving average of the ERM (Fig. \ref{daily_clim_and_era_rf}c) shows that it has doubled from one to two during this period.

\begin{figure*}
    \centering
    \includegraphics[width=1\linewidth]{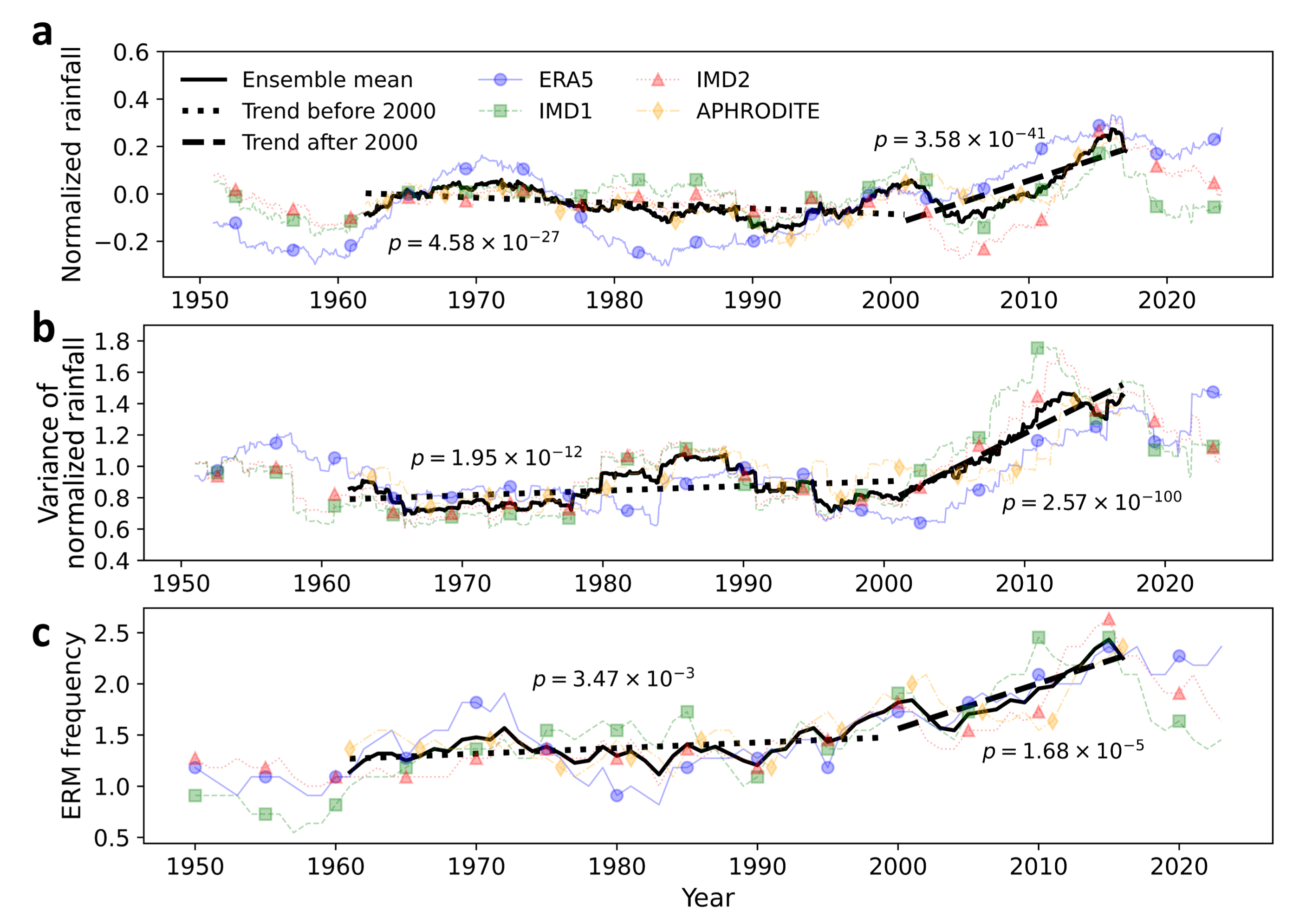}
    \caption{(a) 11-year running mean of monthly mean rainfall anomalies for different datasets and their ensemble mean. (b) Variance of monthly rainfall over 11-year running windows for different datasets and their ensemble mean. (c) 11-year running mean of ERM frequency for different datasets and their ensemble mean. Trend lines for ensemble mean before and after 2000 highlight the recent increase in monthly rainfall, variability and ERM counts over Tamil Nadu. In plots (a) and (b), 120 data points occur within a 10-year interval on the abscissa (x-axis), although each point represents an average over 132 months (11 years). In plot (c), 10 data points occur within a 10-year interval, with each point representing an average over 11 years.
    }
    \label{running_means_rf}
\end{figure*}

To establish the robustness of the increasing trend of the monthly mean rainfall and its variability in recent decades indicated by ERA5 data (Fig. \ref{daily_clim_and_era_rf}b), we compare an 11-year moving average of the monthly rainfall from four different datasets, namely ERA5, IMD1, IMD2 and APHRODITE. Their corresponding ensemble mean for the common time period based on the data availability from these various sources (1951-2015) is shown in Fig. \ref{running_means_rf}a. All the moving averages of monthly rainfall shown are normalized by SD. However, the SD of the original total monthly rainfall is 9.98 m for ERA5, 0.99 m for IMD1, 9.57 m for IMD2, and 9.20 m for APHRODITE. The original total monthly rainfall is calculated from spatial and temporal accumulation of daily rainfall data, and the differences in the SD values reflect the differences in the spatial resolution of the datasets. While the trend of monthly rainfall over Tamil Nadu remains nearly unchanged from 1950 till about 2000 (Fig. \ref{running_means_rf}a), it shows a significant (p$<$0.05) increasing trend thereafter. Time series of monthly mean rainfall over Tamil Nadu from ERA5 (Fig. \ref{daily_clim_and_era_rf}b) also indicates a potential increase in interannual variance. To quantify it, the variance of monthly rainfall over 11-year moving windows is calculated from all datasets and their ensemble mean is shown in Fig. \ref{running_means_rf}b. Notably, the inter-annual variability of the monthly mean has a statistically significant increasing trend in recent decades, after the year 2000. It is notable that the variance of monthly rainfall over Tamil Nadu has more than doubled from the pre-2000s to the present day. Thus, during the past two decades, both mean rainfall and its inter-annual variability have been increasing, making the prediction of monthly rainfall more difficult in recent years.

Finally, in Fig. \ref{running_means_rf}c, the ensemble mean of the 11-year moving average of ERM frequency also shows a statistically significant increasing trend. In all the above cases of rainfall, rainfall variability, and ERM frequency, a broadly similar trend is observed for other datasets. However, the magnitude of these trends differs among datasets, highlighting inherent differences arising from gauge density, interpolation techniques, and reanalysis data assimilation methods \citep{sun2018review, tarek2021uncertainty}. (The statistical significance of the trend lines for individual datasets is presented in Table S1 of the supplementary material).

\subsection{Drivers of rainfall variability over Tamil Nadu} \label{Drivers of rainfall variability over Tamil Nadu}

\begin{figure}
    \centering
    \includegraphics[width=1\linewidth]{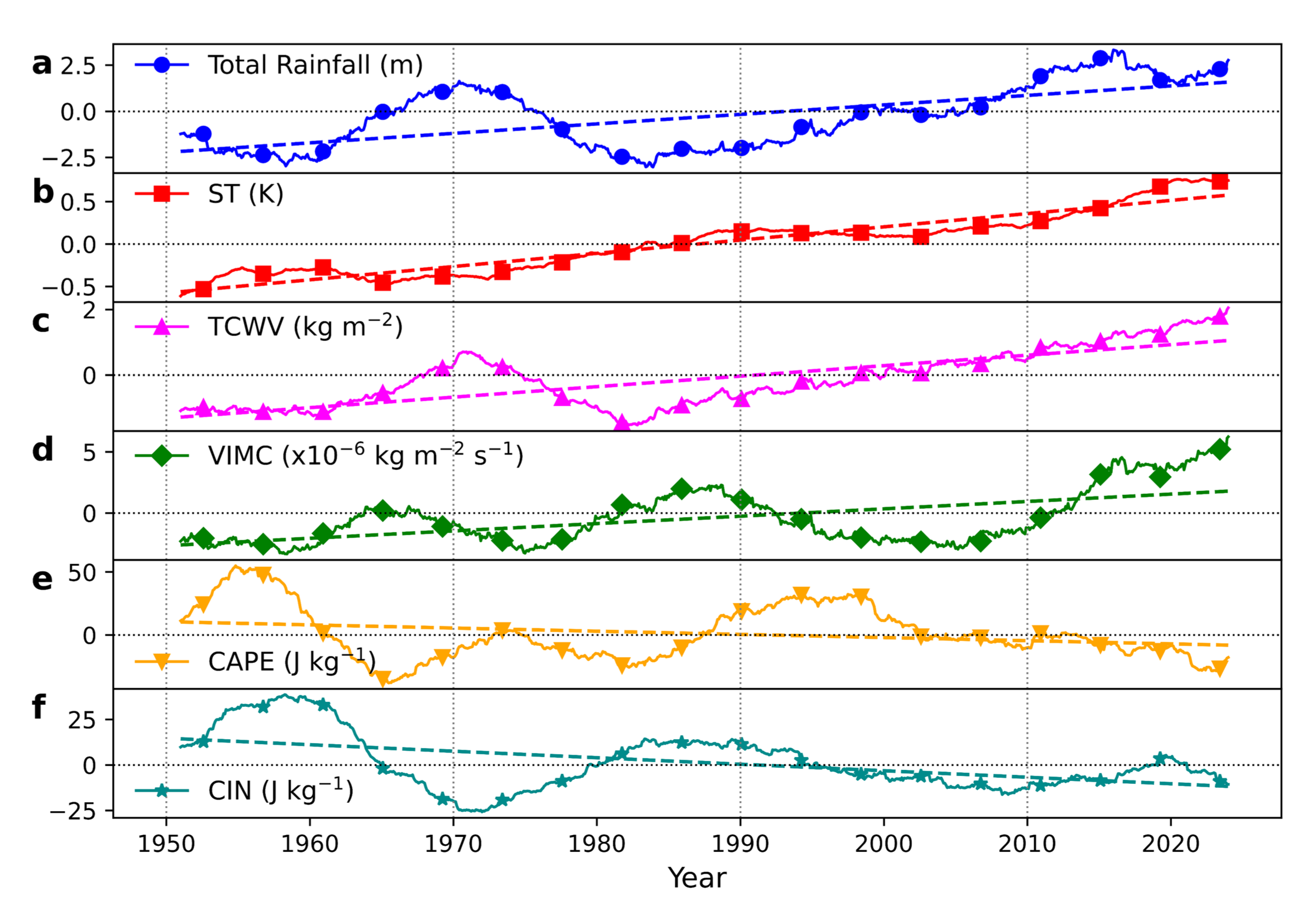}
    \caption{(a-f) 11-year running means of anomalies of total monthly rainfall, surface temperature (ST), total column water vapour (TCWV), vertical integral of moisture convergence flux (VIMC), convective available potential energy (CAPE) and convective inhibition (CIN), all derived from ERA5.
}
    \label{running_means_all_variables}
\end{figure}

What drives the increasing trend of the mean, its inter-annual variability and ERM frequency over Tamil Nadu? Rainfall variability over a region is controlled not only by the local precipitation processes but also by the thermodynamic state of the atmosphere and the large-scale dynamical conditions that regulate moisture supply and convection \citep{trenberth1999conceptual, trenberth2003changing, goswami2006increasing, o2015precipitation}. Therefore, it is useful to analyse key atmospheric variables that directly influence precipitation generation. Surface temperature (ST) and total column water vapour (TCWV) describe the thermodynamic environment and moisture availability, while vertically integrated moisture convergence (VIMC) represents the dynamical transport and accumulation of moisture \citep{trenberth2011changes, roxy2017threefold, chansaengkrachang2018vertically, sengupta2023seasonal}. In addition, convective available potential energy (CAPE) and convective inhibition (CIN) characterize the stability of the atmosphere and the conditions governing the initiation of deep convection \citep{riemann2009global, chen2020changes}. Together, these variables provide a physically consistent framework to understand the mechanisms underlying rainfall changes.

To understand the evolution of rainfall over Tamil Nadu, we complement the 11-year running mean rainfall anomalies with the corresponding 11-year running means of anomalies in ST, TCWV, VIMC, CAPE, and CIN (Fig. \ref{running_means_all_variables}). The results indicate that the recent increase in rainfall over the Tamil Nadu region is associated with enhanced atmospheric moisture availability and favourable dynamical conditions. Figures \ref{running_means_all_variables}b–d show that ST, TCWV, and VIMC have increased in recent decades. 

The rise in ST (Fig. \ref{running_means_all_variables}b), along with its increasing trend, enhances the moisture-holding capacity of the atmosphere, leading to higher TCWV (Fig. \ref{running_means_all_variables}c), which also exhibits a robust increasing trend. This behaviour is consistent with Clausius–Clapeyron scaling, which predicts an approximate increase of atmospheric moisture with warming \citep{trenberth2011changes, sengupta2023seasonal}. This increase in atmospheric moisture provides a larger reservoir of water vapour available for precipitation. At the same time, the strengthening trend of VIMC, as shown in Fig. \ref{running_means_all_variables}d, indicates an increased dynamical transport and accumulation of moisture over the region \citep{chansaengkrachang2018vertically}. Enhanced moisture convergence is a key dynamical mechanism that supports sustained precipitation by continuously replenishing atmospheric moisture.

In contrast, the behaviour of CAPE and CIN (Fig. \ref{running_means_all_variables}e and \ref{running_means_all_variables}f), together with their long-term trends, suggests that the rainfall increase is not primarily associated with stronger deep convective instability. CAPE shows a gradual decrease in recent decades, with a decreasing trend in the long-term. While CIN also exhibits a decreasing trend overall, despite a slight increase in the most recent years. CAPE represents the amount of buoyant energy available to drive deep convection, whereas CIN quantifies the energy barrier that must be overcome for convection to initiate \citep{riemann2009global, chen2020changes}. The decreasing trend in CAPE, therefore, implies that the observed increase in rainfall is not driven by stronger convective instability. However, the relatively weak CIN indicates that the barrier to convection remains weak, allowing convection to be triggered more easily under favourable moisture conditions. This is further supported by the weakening trend in vertical velocity, indicating reduced subsidence over Tamil Nadu. The running mean of vertical velocity at 500 hPa (Fig. S1) exhibits a notable decreasing trend, where the negative values indicate upward motion. This suggests strengthening of mid-tropospheric ascent over the region, consistent with the decreasing CIN.  

Therefore, the recent increase in rainfall over the Tamil Nadu region is primarily linked to thermodynamic moistening of the atmosphere and enhanced dynamical moisture convergence, rather than a strengthening of convective instability. The combination of higher atmospheric moisture content, increased moisture convergence, and relatively weak convective inhibition creates favourable conditions for precipitation. In addition, even though the overall available energy for convection (CAPE) decreases, the cap on CAPE due to CIN is also lower. This allows convection to be triggered more often, leading to more frequent rainfall. Together, these factors contribute to the observed increase in rainfall in recent decades.

\subsection{Changes in the monsoon season} \label{Changes in the monsoon rainfall seasons}

\begin{figure*}
    \centering
    \includegraphics[width=0.8\linewidth]{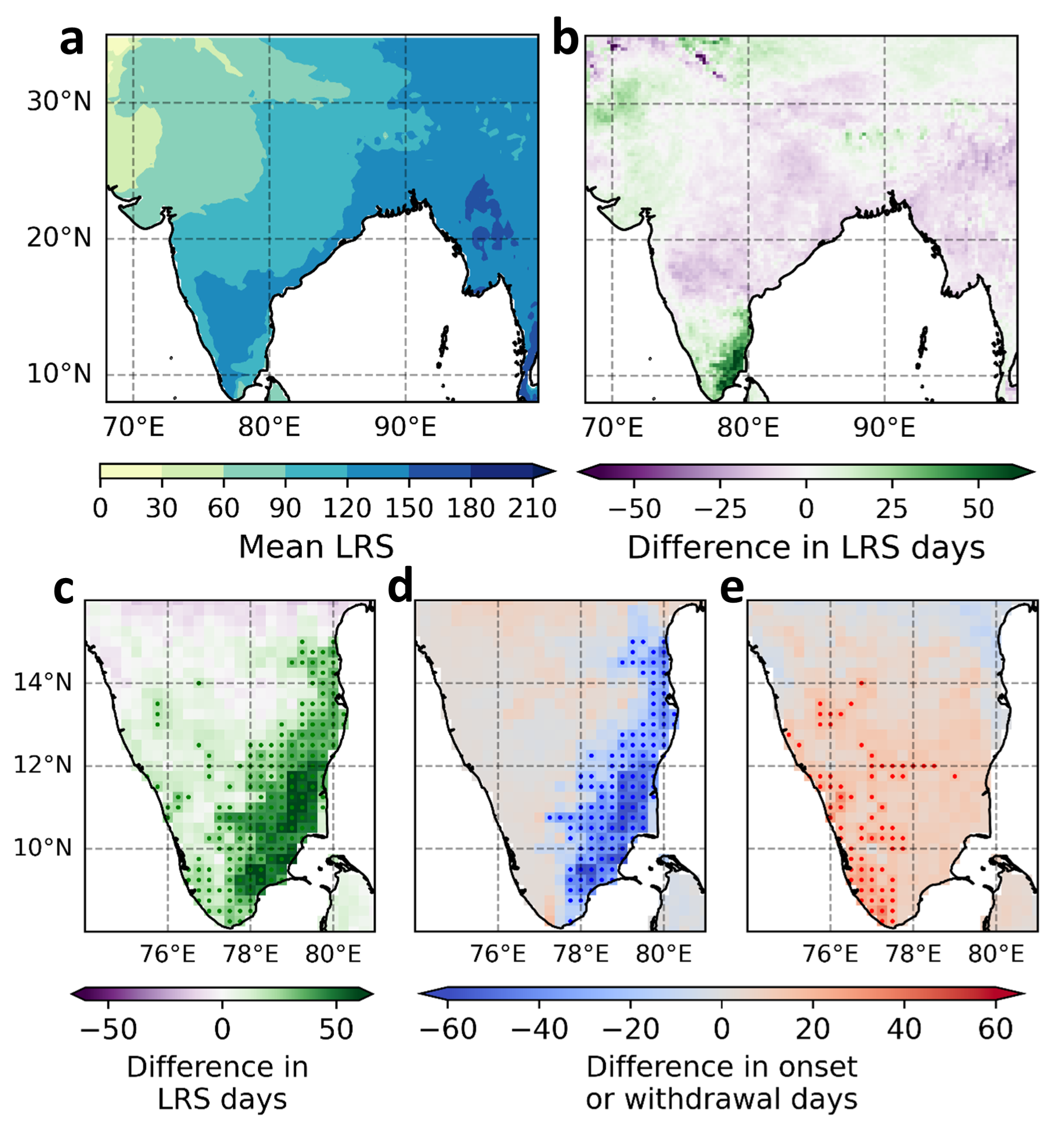}
    \caption{(a) The mean values of the length of the rainfall season (LRS) from 1940 to 2023 are plotted for every location based on the large-scale method. (b) The difference in mean LRS values between time period 1 (1940 - 1960) and time period 2 (2003 - 2023). (c) Similar to (b), the plot focuses on South India, where solid green points are marked to indicate a significant increase in mean LRS by more than 15 days. (d) Difference in mean onset dates between time period 1 and 2, where blue points indicate an advance in onset dates of more than 15 days. (e) Difference in mean withdrawal dates between time period 1 and 2, where red points indicate a delay in withdrawal dates of more than 15 days.
}
    \label{large_scale_monsoon}
\end{figure*}

Figure \ref{large_scale_monsoon}a presents the spatial distribution of the average LRS across India (using the large-scale definition of onset and withdrawal by \cite{patil2025climatic}). Longer rainfall seasons are seen over the southern peninsular regions (including Tamil Nadu), the southeast coastal parts and northeastern India (Fig. \ref{large_scale_monsoon}a). Figure \ref{large_scale_monsoon}b shows the differences in average LRS between period 1 (1940-1960) and period 2 (2003-2023), depicting insight into how LRS over India has changed from the initial decades to the recent decades. The difference between the mean LRS of period 1 and period 2 is plotted at each location on the map. Positive values (green) indicate locations where LRS increased in period 2 compared to time period 1, and negative values (violet) indicate locations where LRS has decreased in period 2. It is noticeable that LRS has increased over the coastal regions of Tamil Nadu and Andhra Pradesh. In Fig. \ref{large_scale_monsoon}c (over South India), every location that undergoes an LRS increase of more than 15 days is highlighted by dark green points, signifying a considerable change in the behaviour of the monsoon.

Figures \ref{large_scale_monsoon}c-e reveal that the LRS increase over Tamil Nadu and Andhra Pradesh is attributed to the advancing monsoon season over those regions (blue points in Fig. \ref{large_scale_monsoon}d), while the LRS increase over some parts of Kerala and Karnataka is attributed to the delayed withdrawal of monsoons (red points in Fig. \ref{large_scale_monsoon}e). The increase in LRS over Tamil Nadu increases the rainfall occurrence during the monsoon season, which contributes to the increasing rainfall and variability observed in the region (Fig. \ref{running_means_rf}). These changes collectively indicate a shift in the rainfall seasonality over southern India.

\subsection{Potential drivers of OND rainfall over Tamil Nadu} \label{Rainfall teleconnections with SST anomalies}

\begin{figure*}
    \centering
    \includegraphics[width=0.9\linewidth]{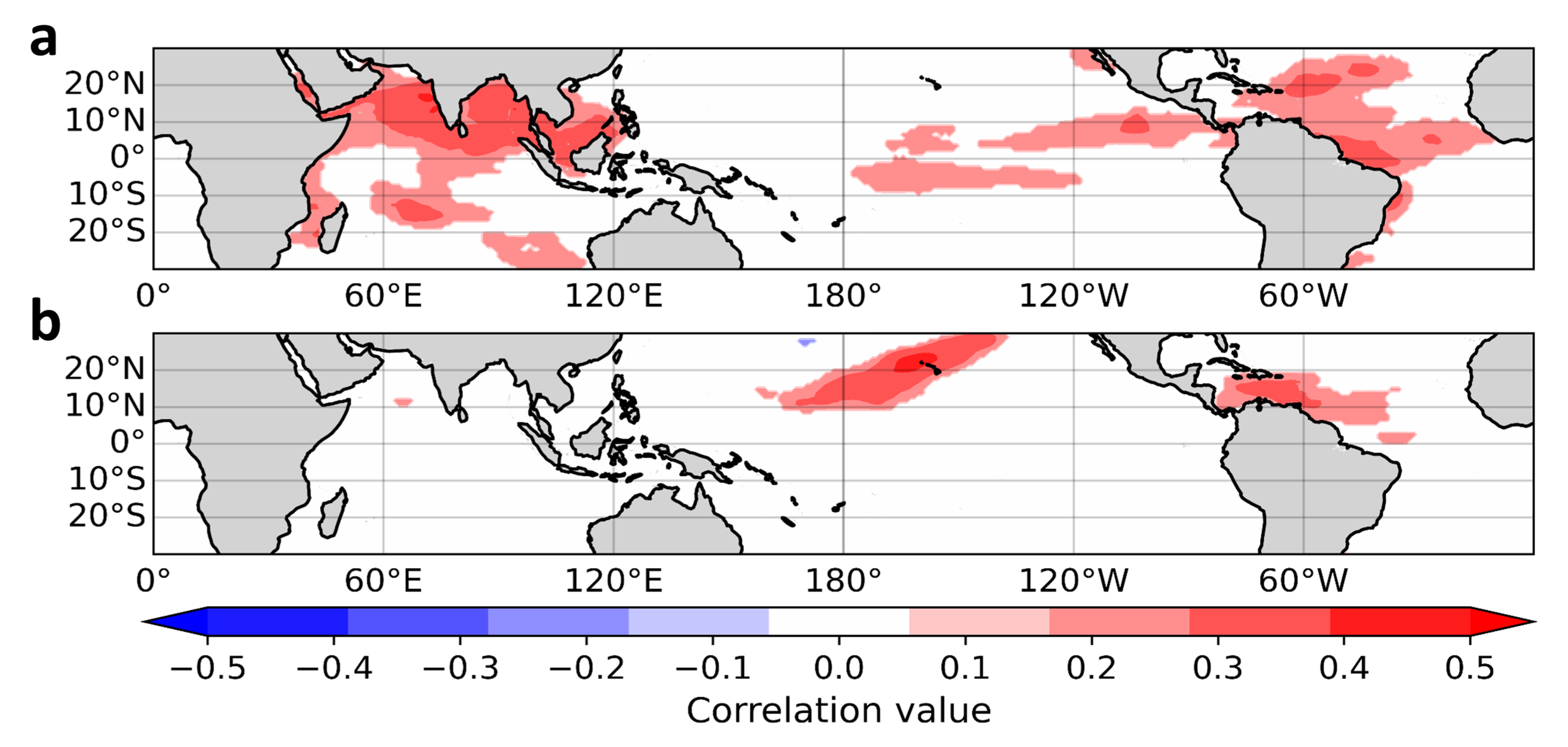}
    \caption{(a) Simultaneous correlation of total OND rainfall (from 1940 to 2023) over Tamil Nadu with global OND average SST anomalies. (b) Correlation between OND rainfall anomalies and SST anomalies at a 10-month lead.
}
    \label{ond_corr_maps}
\end{figure*}

The daily climatology of annual rainfall over Tamil Nadu derived from multiple datasets (Fig. \ref{daily_clim_and_era_rf}a) shows that the largest fraction of annual rainfall occurs during the OND season. Consequently, interannual variability in annual rainfall is largely governed by the OND rainfall variability. Despite the significant variability in recent years, long-lead seasonal prediction of OND rainfall over Tamil Nadu has received little attention. In this section, we try to identify the key drivers of OND rainfall that are essential for developing long-lead seasonal forecasts.

Evaporation being a nonlinear function of SST, small changes in SST can lead to large changes in evaporation and strongly influence local convective activity in the tropical regions where the mean SST is high. Large patches of warm waters in the tropics, therefore, could enhance large-scale convective activity and influence large-scale circulation and teleconnection. Hence, the association of SST with the OND rainfall over Tamil Nadu may provide insight into potential drivers of rainfall variability over the region. The simultaneous (0-month lead) Pearson correlation between OND rainfall anomalies over Tamil Nadu and SST anomalies in each oceanic grid point (Fig. \ref{ond_corr_maps}a) shows significant positive correlation with SST over the North Indian Ocean and the western Pacific warm pool, Bay of Bengal, Arabian Sea and South China Sea, as well as the eastern equatorial Pacific and equatorial Atlantic. These correlations are consistent with results from previous studies and indicate that large-scale oceanic variability influences OND rainfall over Tamil Nadu \citep{zubair2006strengthening, sreekala2012northeast, rajeevan2012northeast, yadav2012enso}. However, these relationships are based on linear correlations and may not adequately capture the complex interactions influencing the rainfall over Tamil Nadu \citep{gadgil2006asian, rajeevan2007new}. Moreover, when this relationship is examined at longer lead times using linear correlation, the spatial patterns become weak and poorly organized (Fig. \ref{ond_corr_maps}b), limiting their usefulness for long-lead prediction.

\begin{figure*}
    \centering
    \includegraphics[width=0.9\linewidth]{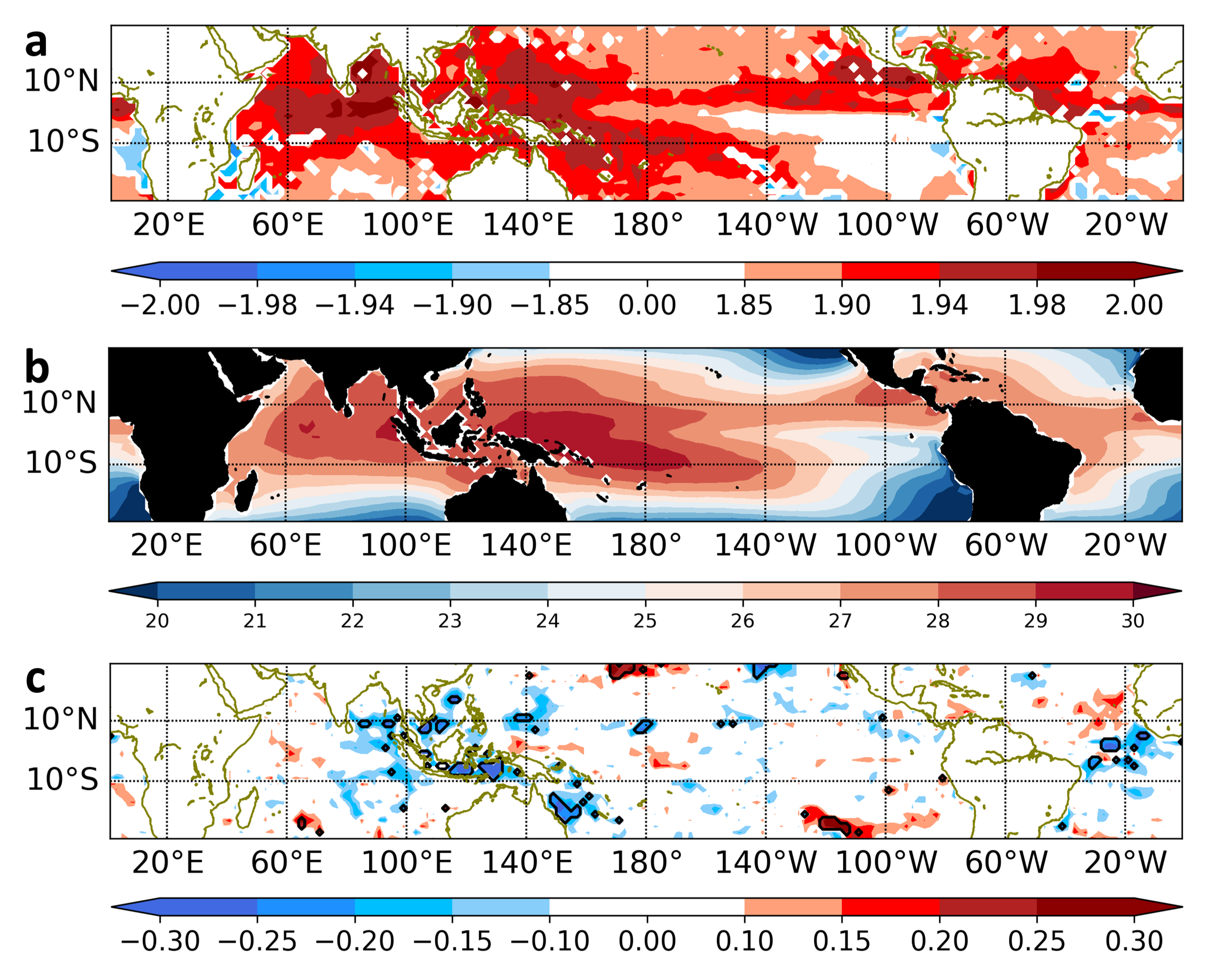}
    \caption{(a) Multi-year mean of the outgoing network strength ($CN^{y}_{i}$) for the period 1941-2023.(b) Annual mean climatology of SST between 1941 and 2023. (c) Pearson correlation between outgoing network strength (1941-2023) and OND rainfall anomalies (1942-2024). Black contours denote regions where the correlation is statistically significant at the 95\% confidence level based on the p-value statistic.}
    \label{CN_with_ sig_test_plots}
\end{figure*}

To overcome this hurdle, we adapt and extend the complex network framework for predictor discovery of seasonal monsoon rainfall similar to that of \cite{ran2025tropical} using daily SST anomalies. In our framework, each grid box represents a network node, and the node-wise network metric quantifies the influence exerted by the SST variability at that location on the SST variability elsewhere, characterized by the outgoing network strength. Physically, nodes with large outgoing strength correspond to SST regions where daily anomalies repeatedly lead SST anomalies at many other locations, with large-magnitude lagged correlations persisting across a wide range of time lags within the year. Although the network is derived from daily SST anomalies, the resulting yearly climate network maps do not represent short-term or weather-scale variability. Random, high-frequency SST anomalies do not contribute systematically to the outgoing strength, whereas SST anomalies associated with slowly evolving ocean–atmosphere coupled processes generate consistent lead–lag relationships that dominate the network structure. In this manner, daily SST variability is effectively translated into information relevant at seasonal time scales. This is evident from the spatial pattern obtained from the multi-year mean of the yearly network maps (Fig. \ref{CN_with_ sig_test_plots}a). Figure \ref{CN_with_ sig_test_plots}a highlights well-known centres of tropical climate interaction such as the Indo-Pacific warm pool and the Atlantic warm pool that can affect tropical convection and large-scale atmospheric circulation \citep{wang2008decadal, de2016indo, park2019effect}. This is evident from the annual mean climatology of the SST between 1941 and 2023 (Fig. \ref{CN_with_ sig_test_plots}b). 

As mentioned in Section \ref{Network Construction}, by construction, the correlation represents the relationship between the year-wise outgoing network at each grid point for year $y$ and the OND rainfall anomalies over Tamil Nadu for year $y+1$, with a lead of 10 months. The correlation between climate network maps and OND rainfall at a 10-month lead exhibits statistically significant signals not only locally but also in remote warm tropical regions, including the South China Sea, the Pacific warm pool, the equatorial Atlantic, and off-equatorial Indian and Pacific Oceans (Fig. \ref{CN_with_ sig_test_plots}c). These relationships are not apparent in direct SST–rainfall correlations at this lead time (Fig. \ref{ond_corr_maps}b) and have not been previously reported. However, previous studies have shown that atmospheric circulation features such as outgoing long-wave radiation (OLR), mean sea level pressure, and Walker circulation over the South China Sea and Pacific warm pool are linked to excess and deficient OND rainfall during the season itself \citep{zubair2006strengthening, sreekala2012northeast, yadav2012enso}. The consistency of these regions with known seasonal circulation influences suggests that the network-based analysis identifies physically plausible regions that may modulate OND rainfall over Tamil Nadu, thereby providing precursors at long-lead times. Hence, using such a network-based encoding climate interactions to make predictions can be promising.

\subsection{Predictability of OND rainfall of Tamil Nadu} \label{Predictability of OND rainfall of Tamil Nadu}

Here, we demonstrate the feasibility of forecasting OND seasonal rainfall over Tamil Nadu based on the newly identified network predictor nodes at a 10-month lead. By following the steps in section \ref{Predictor Selection}, only those nodes that exhibit both significant linear correlation and significant phase synchrony with the target OND rainfall anomalies are retained.

\subsubsection{Potential Predictability Estimate} \label{Potential Predictability Estimate}

Potential predictability refers to the achievable skill in forecasting a climate variable like rainfall using the predictor nodes \citep{saha2016potential, sharma2022mechanism}. The selected predictor nodes for the period 1941-2023, together with OND rainfall anomalies for 1942-2024, are used to develop a ridge regression model to estimate the potential predictability of OND rainfall anomalies over the target period 1982-2024. An out-of-sample (leave-one-year-out) strategy is adopted, in which the model is trained separately for each target year by excluding the predictor nodes corresponding to the year preceding the target and the OND rainfall anomaly of the target year itself. For example, to forecast the OND rainfall anomaly of 1982, predictor nodes from 1981 and the OND rainfall anomaly of 1982 are excluded from the training data set. The trained model is then applied to the predictor nodes from 1981, and the resulting estimate is compared with the observed OND rainfall anomaly for 1982. This procedure is repeated for all target years.

\begin{figure*}
    \centering
    \includegraphics[width=1\linewidth]{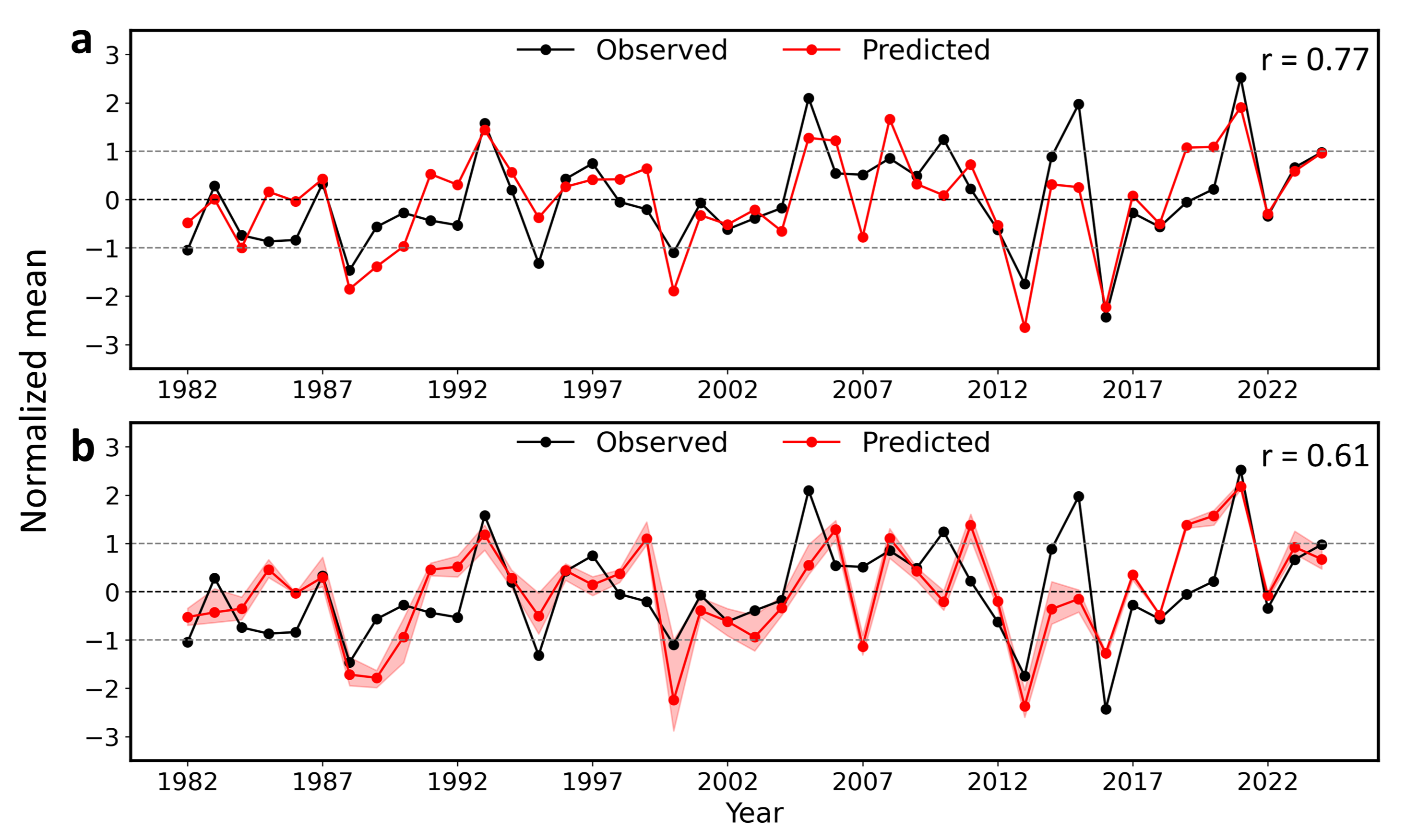}
    \caption{(a) Potential predictability estimate of OND rainfall anomaly between 1982 and 2024. (b) Ensemble mean (red line) and min-max spread (red shading) of 10 independent forecasts of OND rainfall anomaly between 1982 and 2024. The observed OND rainfall anomaly is shown in black. The correlation skill (\textit{r}) is indicated.
}
    \label{predictability}
\end{figure*}

Figure \ref{predictability}a shows the forecast of OND rainfall anomalies over Tamil Nadu for the period 1982-2024. Both the observed and predicted OND rainfall anomalies are standardized by dividing by their respective standard deviations. The correlation skill between the observed and predicted anomalies is 0.77, which is a promising result indicating for the first time the existence of the potential for skillful seasonal forecast for the OND rainfall of Tamil Nadu. Considering normalized OND rainfall anomalies between -1 and +1 are classified as normal, values below -1 as deficient, and values above +1 as excess. We find that 3 out of 4 excess years are successfully predicted, while 4 out of 5 deficient years are correctly captured by the model (Fig. \ref{predictability}a). Because the model training excludes the target year, whereas predictor node selection is performed using the full data set, the resulting correlation skill of 0.77 represents an estimate of potential predictability rather than independent forecast skill. The high potential predictability at a 10-month lead time, together with the model’s ability to capture most excess and deficient events, highlights the physical relevance and predictive value of the OND rainfall precursors derived from the complex climate network.

\subsubsection{Independent Forecast Verification} \label{Independent Forecast Verification}

For independent forecast verification, both the predictor node selection and the model training are performed using a fully out-of-sample strategy by excluding the OND rainfall anomaly of the target year and the corresponding predictor nodes. This contrasts with the potential predictability estimate, in which only the model training excludes the target year, while the predictor node selection is performed using the full data set. 

An ensemble of 10 independent forecasts is generated. The ensemble approach arises from the surrogate-based significance testing used to identify predictor nodes, in which randomized surrogate data are employed to estimate the statistical significance of the network-rainfall relationships. The stochastic nature of the surrogate procedure allows multiple realizations of predictor selection and, consequently, multiple independent forecasts. Figure \ref{predictability}b shows the ensemble-mean prediction along with the ensemble spread. Both the observed and predicted OND rainfall anomalies are standardized over the target period 1982-2024. The independent forecast exhibits a correlation skill of 0.61 at a 10-month lead. Based on categorical evaluation, 2 out of 4 excess rainfall years and 4 out of 5 deficient rainfall years are correctly predicted.

\begin{table}
\begin{tabular*}{\textwidth}{@{\extracolsep\fill}lccccccccc}
\toprule%

& & & \multicolumn{3}{c}{True Positive Rate} & \multicolumn{3}{c}{False Positive Rate} & \\\cmidrule(lr){4-6}\cmidrule(lr){7-9}%

r & HSS & Accuracy & Deficient & Normal & Excess & Deficient & Normal & Excess &MSSS    \\

\midrule
0.61  & 0.34 & 0.70 & 0.67 & 0.75 & 0.40 & 0.05 & 0.46 & 0.16 & 0.21\\

\botrule
\end{tabular*}
\caption{Categorical forecast verification} \label{pred_table}
\end{table}

The performance of the independent forecast is further evaluated using multiple categorical and probabilistic verification metrics (Table \ref{pred_table}). The Heidke Skill Scores (HSS = 0.34), which evaluate forecast skill against random chance, indicate substantial improvement over random guess for categorical rainfall prediction, demonstrating that the model possesses skill in separating deficient, normal, and excess rainfall years. The HSS ranges from $-\infty$ to 1, with zero indicating no skill relative to random chance. The overall categorical accuracy of 0.70 shows that 70\% of the years are correctly classified into the three rainfall categories. Category-wise hit rates, quantified by the True Positive Rate (TPR), reveal that the model performs best for deficient years (TPR = 0.67), followed by normal years (TPR = 0.75), while skill for excess years is comparatively lower (TPR = 0.40). The Mean Squared Skill Score (MSSS = 0.21), computed relative to a climatological reference forecast, also ranging from $-\infty$ to 1 and positive only when the forecast outperforms climatology, demonstrates that the predictions are meaningfully better than the climatological guess.

Achieving a correlation skill of 0.61 at a 10-month lead represents a substantial advancement, as comparable long-lead forecast skill for OND rainfall over Tamil Nadu is not available from existing statistical or dynamical systems. This further underscores the relevance of the network-based predictors. Notably, both the potential predictability and independent forecast results indicate higher skill in predicting deficient rainfall years than excess rainfall years. This asymmetry may reflect stronger event-to-event variability in the processes governing excess rainfall, which are not fully captured by the present regression framework. Given that the potential predictability reaches 0.77, there remains considerable scope for improving independent forecast skill using more advanced modelling approaches that can better represent such relationships.


\section{Discussions}\label{Discussions}

While the coefficient of variability of seasonal rainfall over India at the all-India scale (ISMR) is 10\% of the long-term mean, that over smaller spatial scales, such as the meteorological subdivisions, is 25\% of the long-term mean during the JJAS season, while as large as 100\% in the Northwest India and about 40\% in the Southeast Tamil Nadu (Chauhan et al., 2022). Hence, the prediction of sub-regional seasonal rainfall, such as that over Tamil Nadu, is intrinsically more difficult than that of all-India monsoon rainfall, and the limit on deterministic predictability would be shorter than that for the all-India rainfall. However, from the perspective of users, the sub-regional rainfall prediction, such as that over Tamil Nadu, is more useful than the prediction of all-India monsoon rainfall. Tamil Nadu receives the highest rainfall in the northern winter (OND) season. The basis for long-lead predictability of ISMR is established in \cite{sharma2022mechanism}, and a physics-guided AI/ML model can help realize the potential predictability of ISMR \citep{sharma2026improving}. Neither such an estimate of the long-lead predictability of OND rainfall over Tamil Nadu nor a model to achieve that is currently available. This problem is not just the problem of Tamil Nadu, but a more generic problem of prediction and predictability of sub-regional rainfall. To our best knowledge, no such estimate of long-lead predictability for any sub-regions smaller than India (e.g., the four homogeneous regions, \cite{kothawale2017rainfall}) is available. 

To fill this gap, we embark on a study of the variability and predictability of monthly rainfall over Tamil Nadu. Our study provides long-term rainfall assessment by utilizing multiple rainfall datasets and the atmospheric factors influencing the Tamil Nadu region from 1940 to 2023. An increase in the frequency of extreme rainfall months and rainfall variability over Tamil Nadu in recent decades highlights the increasing frequency of rainfall and wetter conditions over the region. This behavior is consistent with rising surface temperature, increased total column water vapour, and strengthening moisture convergence, all of which are in favor of enhanced precipitation. The analysis of convective parameters reveals a shift in the drivers of rainfall. While convective available potential energy shows a decreasing trend, indicating weaker potential for strong convection, the convective inhibition values have reduced in recent decades. Together, these factors support a conclusion that the large-scale changes in circulation over the region have been helping stronger moisture convergence and higher rainfall despite the reduced convective available potential energy. This explains the observed rise in the frequency of extreme rainfall months as well.

The seasonal rainfall climatology shows that Tamil Nadu receives significant rainfall during OND. We observe that the monsoon rainfall season exhibits a long-term shift. Compared with the earlier period (1940–1960), the recent period (2003–2023) shows an increase in the length of the rainfall season over southern peninsular India. This increase in the length of the rainfall season results from earlier onset dates over Tamil Nadu and Andhra Pradesh, and delayed withdrawal dates over Kerala and Karnataka. Exploring the causes of the rapid increase in the mean and variability of monthly rainfall over the past two decades remains an important area for future work.

Following these leads, we use complex climate networks of SST to unravel an SST-based predictor of OND rainfall over Tamil Nadu. The analysis reveals that the changes in SST anomalies in some oceanic regions precede OND rainfall by 10 months, indicating a source of potential predictability. By incorporating directionality in the climate network framework, we are able to distinguish SST regions that act as drivers or precursors for predicting OND rainfall. The high predictability at a 10-month lead time and the ability to capture most of the deficient, normal and excess OND seasonal rainfall over Tamil Nadu demonstrate that the SST precursors derived from the climate network contain physically meaningful and independently predictive information, even in the presence of increasing variability of rainfall. In addition, this study establishes the feasibility of estimating long-lead predictability and independent out-of-sample forecast skill for seasonal rainfall using network-derived predictors.

An estimate of the growth of errors based on a 5-month moving average of the all-India rainfall time series (Fig. S4a) indicates that the deterministic limit of predictability of all-India rainfall is about 6 months (Fig. S4b). Thus, the limit of deterministic predictability of Tamil Nadu rainfall should be even shorter, because of its high coefficient of variability. In that backdrop, our finding of high potential skill ($r$ = 0.77) of OND rainfall over Tamil Nadu may be considered non-intuitive. However, nonlinear lag synchronization between two interacting chaotic oscillators \citep{pecora1990synchronization, rosenblum1996phase, rosenblum1997phase} (e.g. OND rainfall over Tamil Nadu and 10-month lead predictor) provides the basis for long-lead predictability in several physical systems. 

Lag synchronization occurs when the trajectories of two dynamical systems evolve nearly identically, but are separated by a finite time delay, while maintaining a consistent phase relationship between their oscillatory components, such that $Y(t)\approx X(t-{\tau})$. Here $X(t)$ and $Y(t)$ represent the two dynamical systems and ${\tau}$ is the time delay. In geophysical systems, such as the climate, lag-synchronization maybe omni present but difficult to extract from observations. The high correlation ($r$ = 0.77) between OND rainfall ($Y(t)$) and the 10-month lead predictor ($X(t)$) is evidence of finding such a lag synchronization from observed geophysical data. Our predictor discovery method has been successful in extracting the same.  Therefore, we argue that there is a physical and dynamical basis for such long-lead predictability far beyond the deterministic limit on predictability. One limitation of our predictor discovery method, however, is that for the OND season, the minimum lead time for which we could construct the predictor is a 10-month lead, and it is not possible to build predictors for all lead months from 1-month to say 24-months and find the lead month for which the potential predictability is highest. We plan to overcome this limitation in future work.

Our findings indicate that Tamil Nadu is undergoing a gradual shift toward increased rainfall activity, and the development of a successful 10-month lead OND rainfall prediction model for Tamil Nadu has important implications for water resource management, extreme rainfall preparedness, and agricultural planning.


\subsubsection*{Acknowledgments}
The authors are grateful for the funding and support provided by the IndusInd Bank for pursuing this research under grant No. CR23242519AEINIB002696. The authors thank Matilda Lobo, Devraj Yadav, Hemangi Patil and Tejashree Wadivkar from IndusInd Bank for their support and guidance. We also acknowledge the Office of Institutional Advancement and the Alumni and Corporate Relations Office at the Indian Institute of Technology Madras for administrative support. The authors also acknowledge the Institute of Eminence Initiative of the Indian Institute of Technology Madras for access to the high-performance computing (HPC) cluster facilities used in this study (No. SP22231222CPETWOCTSHOC). B. N. Goswami is thankful for support from Anusandhan National Research Foundation through the ANRF Prime Minister Professorship.

\subsubsection*{Author Contributions} 

The study was conceived and designed by Devabrat Sharma, Gaurav Chopra, B. N. Goswami, R. I. Sujith and Harini S. Material preparation, data collection, and visualization were performed by Harini S and Devabrat Sharma. The large-scale monsoon network analysis was carried out by Yogenraj Patil, and its conceptual framework was developed by Gaurav Chopra, Shruti Tandon, B. N. Goswami, and R. I. Sujith. R. I. Sujith supervised the study and arranged funding. The manuscript was written by Harini S and Devabrat Sharma, while B. N. Goswami, Yogenraj Patil, Gaurav Chopra, Shruti Tandon, and R. I. Sujith provided critical feedback and contributed to revisions of the manuscript. All authors read and approved the final manuscript.

\subsubsection*{Funding}
This work was supported by IndusInd Bank under grant No. CR23242519AEINIB002696.

\subsubsection*{Data Availability}
The datasets analyzed during the current study are publicly available from their respective sources. ERA5 reanalysis data from ECMWF can be accessed through the Climate Data Store (\url{https://cds.climate.copernicus.eu}). Gridded rainfall data based on observations from the India Meteorological Department are available in IMD data portals (\url{https://www.imdpune.gov.in/lrfindex.php}; \url{https://www.imdpune.gov.in/cmpg/Griddata/Rainfall_25_NetCDF.html}). APHRODITE precipitation data can be accessed from \url{http://aphrodite.st.hirosaki-u.ac.jp}. COBE-SST2 sea surface temperature data from JMA can be accessed at \url{https://psl.noaa.gov/data/gridded/data.cobe2.html}. The codes used in this study can be provided upon reasonable request.

\subsubsection*{Supplementary Information} 
The online version contains supplementary material, including a detailed description of the methodology.

\section*{Declarations}
\subsubsection*{Competing interests} The authors declare that there are no competing interests.



\clearpage

\begin{center}
\begingroup
\linespread{1.3}\selectfont

{\Large \textit{Supplementary materials for}}\\[6pt]

{\huge Prediction and Predictability of the Wet-Season Rainfall over Southeast India}\\[10pt]

{\large
Harini S$^{1,2}$, Devabrat Sharma$^{1,2*}$, Yogenraj Patil$^{1,2}$,\\
Gaurav Chopra$^{3}$, Shruti Tandon$^{1,2}$, B. N. Goswami$^{4}$,\\
R. I. Sujith$^{1,2*}$\\[6pt]

$^{1}$Department of Aerospace Engineering, Indian Institute of Technology Madras, Chennai, India\\
$^{2}$Centre for Excellence for studying Critical Transitions in Complex Systems, IIT Madras, Chennai, India\\
$^{3}$Department of Applied Mechanics, IIT Delhi, India\\
$^{4}$ST Radar Centre, Gauhati University, Guwahati, India\\[6pt]

*Corresponding author:\textit{\textcolor{blue}{devabratsharma9597@gmail.com}}

}
\endgroup
\end{center}

\subsubsection*{Contents}
\begin{itemize}
\setlength{\itemsep}{2pt} 
\renewcommand{\labelitemi}{} 

\item Text S1 to S5
\item Figure S1 to S4
\item Table S1
\item References

\end{itemize}

\renewcommand{\thesection}{S\arabic{section}}
\setcounter{section}{0}
\setcounter{subsection}{0}
\setcounter{subsubsection}{0}


\renewcommand{\thefigure}{S\arabic{figure}}
\setcounter{figure}{0} 

\renewcommand{\thetable}{S\arabic{table}}
\setcounter{table}{0} 

\section{p-values of rainfall trends of Tamil Nadu} \label{p-values of rainfall trends of Tamil Nadu}
\begin{table}[h]
\centering
\begin{tabular}{llcc}
\toprule
Variable & Dataset & Before 2000 & After 2000 \\
\midrule

Rainfall 
& ERA5       & $1.57\times10^{-1}$ & $4.65\times10^{-62}$ \\
& IMD1       & $3.16\times10^{-14}$ & $4.24\times10^{-1}$ \\
& IMD2       & $6.46\times10^{-1}$ & $7.4\times10^{-40}$ \\
& APHRODITE  & $9.19\times10^{-55}$ & $1.00\times10^{-18}$ \\

\midrule

Rainfall variability 
& ERA5       & $4.27\times10^{-55}$ & $3.24\times10^{-115}$ \\
& IMD1       & $8.96\times10^{-1}$ & $2.1\times10^{-17}$ \\
& IMD2       & $6.31\times10^{-2}$ & $4.83\times10^{-3}$ \\
& APHRODITE  & $1.85\times10^{-27}$ & $1.32\times10^{-45}$ \\

\midrule

ERM frequency 
& ERA5       & $3.14\times10^{-1}$ & $8.73\times10^{-1}$ \\
& IMD1       & $7.55\times10^{-1}$ & $-1.47\times10^{-1}$ \\
& IMD2       & $7.1\times10^{-1}$ & $5.36\times10^{-1}$ \\
& APHRODITE  & $3.97\times10^{-1}$ & $4.45\times10^{-1}$ \\

\bottomrule
\end{tabular}
\caption{p-values of linear trend lines of rainfall characteristics before and after the year 2000}
\label{pvalue_table}
\end{table}

Table \ref{pvalue_table} presents the p-values associated with the linear trends in rainfall of Tamil Nadu, rainfall variability, and excess rainfall months (ERM) frequency for different datasets, calculated for the time period before and after 2000. p-value of less than 0.05 indicates that the trend is statistically significant at the 95\% confidence level. The results show that rainfall variability exhibits statistically significant trends in all datasets after 2000, indicating a robust increase in variability across datasets. Rainfall trends show mixed significance, with ERA5, IMD2, and APHRODITE datasets exhibiting significant trends after the year 2000, while IMD1 shows a weaker or insignificant trend. On the other hand, the trends in ERM frequency are not statistically significant across different datasets.

\section{Running mean of vertical velocity of Tamil Nadu} \label{Running mean of vertical velocity of Tamil Nadu}
\begin{figure*}[h]
    \centering
    \includegraphics[width=0.9\linewidth]{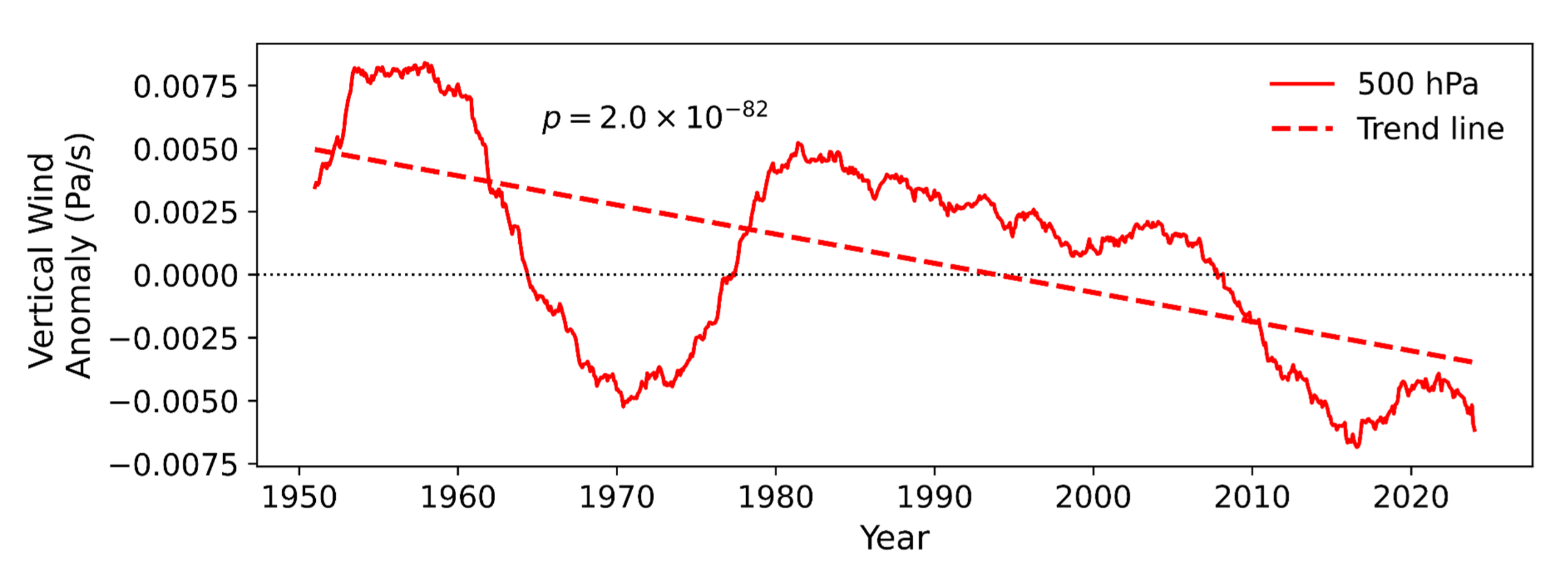}
    \caption{11-year running mean of vertical velocity averaged over Tamil Nadu at 500 hPa, along with its trend.}
    \label{supp_w_wind}
\end{figure*}

The running mean of vertical velocity (w wind; from 1940-2023 derived from ERA5) averaged over Tamil Nadu at 500 hPa is observed to understand the convection and subsidence at mid-troposphere (Fig. \ref{supp_w_wind}). The negative values correspond to the upward motion. The decreasing trend observed during the recent decades suggests a weakening of subsidence and a strengthening of upward motion over Tamil Nadu. This is consistent with the decreasing trend in convective inhibition (CIN) discussed in the main text.

\section{Methodology of Network Construction} \label{Methodology of Network Construction}

A complex network constructed from a gridded climate variable represents the non-trivial statistical interdependence between different spatial locations of the same field. When formulated using time-lagged correlations of the variable across grid points, the network becomes directed and shows how variations at one location tend to influence other locations after some time. We construct a complex network, using daily SST anomalies following the framework of \cite{ran2025tropical}, who developed a similar network based on daily 2m air temperature to generate predictors for forecasting seasonal rainfall over monsoon regions at lead times of 4-10 months. Their linear regression models, constructed using network-derived predictors selected based on their correlation with seasonal monsoon rainfall over different regions, demonstrate high forecast skill, with a correlation coefficient ranging from 0.63 to 0.81. But the predictor selection for model development in Ran et al. (2025) relies on correlation information from both training and testing periods. As a result, reported skill reflects potential predictability rather than an independent out-of-sample forecast verification. Using our framework, we explicitly demonstrate the feasibility of estimating both the long-lead potential predictability and the independent out-of-sample forecast skill for seasonal monsoon rainfall using network-derived predictors.

\subsection{Constructing Adjacency matrices} \label{Constructing Adjacency matrices}
For each year $y$ from 1941 to 2023, a series of complex networks is constructed using time-lagged Pearson correlations of daily SST anomaly over the global tropics ($0^{\circ}$ – $360^{\circ}$E and $30^{\circ}$S - $30^{\circ}$N). Daily SST data for each year are taken from Jan 1 to Dec 31, while lagged correlations are allowed to extend into year $y-1$ to account for delayed interactions. Each $2^{\circ} \times 2^{\circ}$ grid box is treated as a node in the network, resulting in a total of N = 5400 nodes. We choose SST as the network variable because it evolves more slowly than atmospheric fields, which are strongly influenced by stochastic processes. This relatively slow temporal variability makes SST particularly suitable for identifying long-lead precursors relevant for seasonal rainfall prediction.

For a given year $y$, Pearson correlations are computed between all possible pairs of nodes $i$ and $j$ for time lags ranging from 0 to 200 days. The maximum lag of 200 days is chosen to ensure a robust estimation of the background noise level. This procedure yields 201 adjacency matrices per year, each of dimensions 5400 x 5400, with each matrix corresponding to a specific time lag. These matrices are denoted as $A^{y}_{i,j}(\tau)$.

\begin{equation}
{A^{y}_{i,j}(\tau)} = 
\begin{bmatrix}
S^{y}_{1,1}(\tau) & S^{y}_{1,2}(\tau) & \cdots & S^{y}_{1,N}(\tau) \\
S^{y}_{2,1}(\tau) & S^{y}_{2,2}(\tau) & \cdots & S^{y}_{2,N}(\tau) \\
\vdots            & \vdots            & \ddots & \vdots            \\
S^{y}_{N,1}(\tau) & S^{y}_{N,2}(\tau) & \cdots & S^{y}_{N,N}(\tau)
\end{bmatrix}
\end{equation}

Each element of the adjacency matrix is defined as:
\begin{equation}
{S^{y}_{i,j}(\tau) = Corr(SST^{y}_{i}(t-\tau), \,SST^{y}_{j}(t))}
\end{equation}
where Corr(.) denotes the Pearson correlation coefficient, and $t$ spans the daily time indices from Jan 1 to Dec 31 (365 days).

With this formulation, the network is directed. A matrix element ${S^{y}_{i,j(\tau)}}$ quantifies the extent to which SST variability at node $i$ precedes similar variability at node $j$ after a time delay $\tau$. Accordingly, outgoing links from a node indicate locations where SST anomalies tend to occur earlier and subsequently influence other regions, while incoming links identify locations whose variability tends to respond later to remote SST signals.

\subsection{Positive and negative adjacency matrices} \label{Positive and negative adjacency matrices}
Next, for each ordered pair of nodes $(i,j)$ we extract:

\begin{enumerate}[1.]
    \item max(${S^{y}_{i,j}(\tau^{+})}$): the maximum positive correlation across all lags and its corresponding time lag $\tau^{+}$.
    \item min (${S^{y}_{i,j}(\tau^{-})}$): the minimum negative correlation across all lags and its corresponding time lag $\tau^{-}$.
\end{enumerate}

If $\tau^{+}$ ($\tau^{-}$) is greater than zero, then based on the nodes pairs associated with max(${S^{y}_{i,j}(\tau^{+})}$) and max(${S^{y}_{i,j}(\tau^{-})}$), this procedure yields two directed, unweighted adjacency matrices for each year y: the positive adjacency matrix ${PA^{y}_{i,j}(\tau)}$ and the negative adjacency matrix ${NA^{y}_{i,j}(\tau)}$. The elements the positive adjacency matrix ${PA^{y}_{i,j}(\tau)}$ (${NA^{y}_{i,j}(\tau)}$) indicate that variability at node $i$ precedes similar (opposite) variability at node $j$. All entries corresponding to $\tau^{+}$ ($\tau^{-}$) not greater than zero are masked to avoid contamination from instantaneous correlations, ensuring directional causality.

\subsection{Weighted positive and negative adjacency matrices} \label{Weighted positive and negative adjacency matrices}
Each non-zero element of ${PA^{y}_{i,j}(\tau)}$ (${NA^{y}_{i,j}(\tau)}$) are then standardized, yielding two 2-D directed and weighted adjacency matrices for each year: the weighted positive 2-D adjacency matrix ${WPA^{y}_{i,j}}$  and the weighted negative 2-D adjacency matrix ${WNA^{y}_{i,j}}$.
\begin{equation}
{WPA}^{y}_{i,j} =
\frac{{S}^{y}_{i,j}(\tau^{+}) - mean\,({S}^{y}_{i,j})}
{std\,({S}^{y}_{i,j}(\tau))}
\end{equation}
\begin{equation}
{WNA}^{y}_{i,j} =
\frac{{S}^{y}_{i,j}(\tau^{-}) - mean\,({S}^{y}_{i,j})} 
{std\,({S}^{y}_{i,j}(\tau))}
\end{equation}

\subsection{Anti-symmetric matrices} \label{Anti-symmetric matrices}
Now, since the time-lag dimension has been collapsed in the weighted positive and negative networks ${WPA^{y}_{i,j}}$ and ${WNA^{y}_{i,j}}$ and, the direction of influence between nodes is inferred using associated anti-symmetric matrices, ${AP^{y}_{i,j}}$ and ${AN^{y}_{i,j}}$, respectively.

For the positive network, the antisymmetric matrix ${AP^{y}_{i,j}}$ is defined as:
\begin{equation}
AP^{y}_{i,j} =
\begin{cases}
1,  & if\,\,\, WPA^{y}_{i,j} > WPA^{y}_{j,i}, \\[6pt]
-1, & if\,\,\, WPA^{y}_{i,j} < WPA^{y}_{j,i}.
\end{cases}    
\end{equation}

Indicating a directed link from node $i$ to node $j$ or from node $j$ to node $i$, respectively. The antisymmetric matrix ${AN^{y}_{i,j}}$ for the negative network is defined analogously using $WNA^{y}_{i,j}$.

\subsection{Positive and negative correlation networks} \label{Positive and negative correlation networks}

In network terminology, the number of links originating from a node is referred to as out-degree, while the number of links pointing toward a node is referred to as in-degree. As the objective of this study is to identify regions that exert influence on other regions, we focus on outgoing connections. In our framework, the out-degree of node $i$ is given by the total number of nodes $j$ for which ${AP^{y}_{i,j}}$ = 1 (or ${AN^{y}_{i,j}}$ = 1) in the positive (negative) correlation networks. Together with the corresponding link ${WPA^{y}_{i,j}}$ (or ${WNA^{y}_{i,j}}$) we compute the average strength of outgoing links for each node, separately for the positive and negative networks, denoted by ${PCN^{out}_{i}(y)}$ and ${NCN^{out}_{i}(y)}$, respectively.

For the positive (negative) correlation network, considering only those links for which ${AP^{y}_{i,j}}$ = 1 (${AN^{y}_{i,j}}$ = 1), the average outgoing link strength of node $i$ is defined as:
\begin{equation}
{PCN^{out}_{i}(y)} = 
\frac{\sum^{N}_{i=1,i \neq j}WPA^{y}_{i,j}}
{P^{i+}_{out}}
\end{equation}
\begin{equation}
{NCN^{out}_{i}(y)} = 
\frac{\sum^{N}_{i=1,i \neq j}WNA^{y}_{i,j}}
{P^{i-}_{out}}
\end{equation}
where ${P^{i+}_{out}}$ (${P^{i-}_{out}}$) denotes the out-degree of node $i$, i.e., the number of nodes j satisfying ${{AP^{y}_{i,j}}}$ = 1, (${{AN^{y}_{i,j}}}$ = 1).

\begin{figure*}
    \centering
    \includegraphics[width=0.9\linewidth]{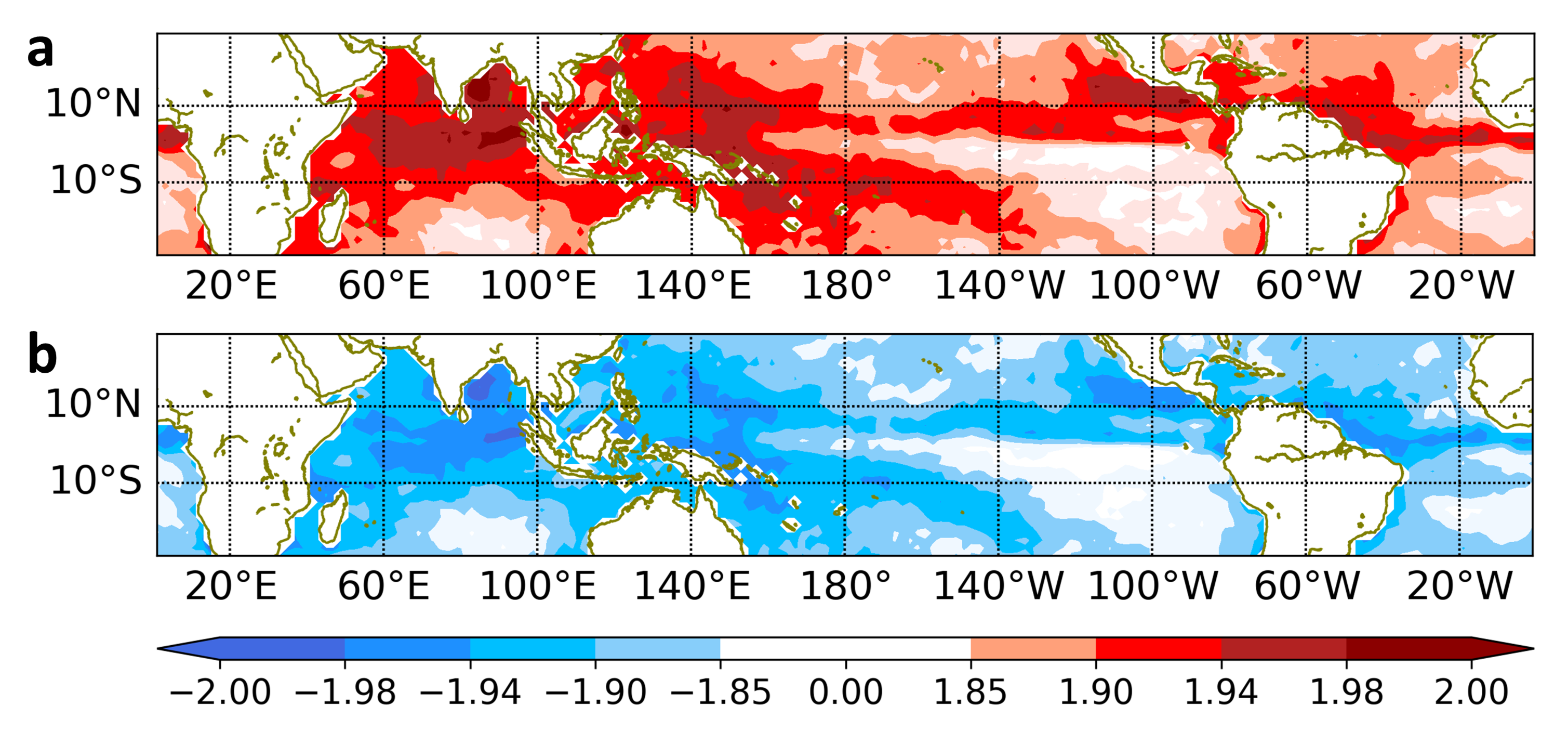}
    \caption{(a) Multi-year mean of the outgoing strength of positive correlation network ($PCN^{y}_{i}$) for the period 1941-2023. (b) Same as (a) but for the negative correlation network ($NCN^{y}_{i}$).
}
    \label{PCN_and_NCN_plots}
\end{figure*}

The node-wise metrics ${PCN^{out}_{i}(y)}$ and ${NCN^{out}_{i}(y)}$ can be mapped back onto the original $2^{\circ} \times 2^{\circ}$ SST grid, yielding two spatial fields of size 30 x 180 for each year. These fields represent the spatial distribution of outgoing influence in the positive and negative correlation networks, respectively (Fig. \ref{PCN_and_NCN_plots}). 

We further combine information from ${PCN^{out}_{i}(y)}$ and ${NCN^{out}_{i}(y)}$ by computing the long-term mean outgoing link strength at each grid box separately for the positive and negative correlation networks. For each year, a single complex network ${CN^{y}_{i}}$ is then constructed by selecting, at each grid box, the network (positive or negative) with the larger mean outgoing strengths.
\begin{equation}
CN^{y}_{i} =
\begin{cases}
PCN^{out}_{i},  & if \, \left|\overline{PCN^{out}_{i}}\right| > \left|\overline{NCN^{out}_{i}}\right|, \\[6pt]
NCN^{out}_{i}, & otherwise.
\end{cases}    
\end{equation}
where $\left|\overline{PCN^{out}_{i}}\right|$ and $\left|\overline{NCN^{out}_{i}}\right|$ are the absolute values of long-term mean of $PCN^{out}_{i}(y)$ and $NCN^{out}_{i}(y)$ over the period 1941-2023, respectively.

\section{Phase Locking value} \label{PLV}

The Phase Locking Value or PLV introduced by \cite{lachaux1999measuring} is a nonlinear metric used to quantify the consistency of phase synchronization between two oscillatory signals over time. Unlike conventional measures, PLV captures the stability of the instantaneous phase relationship independently of signal amplitude and linear dependence.

The PLV ranges from 0 to 1, where a value close to 0 indicates weak or absent phase synchronization, while a value close to 1 represents strong and persistent phase locking between the two signals.

Considering $x(t)$ as a real-valued time series signal, the instantaneous phase of the signal is obtained by first converting it into a complex analytic signal \textit{z}(\textit{t}) using the Hilbert transform:
\begin{equation}
z(t) = x(t) + i\,\,\mathcal{H}[x(t)],
\end{equation}

\begin{equation}
z(t) = A(t)\,e^{i\phi(t)},
\end{equation}
where $A(t)\,e^{i\phi(t)}$ is the polar form of the analytic signal and $\mathcal{H}[x(t)]$ is the Hilbert transform of $x(t)$, with
\begin{equation}
A(t) = |z(t)| = \sqrt{x(t)^2 + (\mathcal{H}[x(t)])^2}
\end{equation}
representing the instantaneous amplitude, and
\begin{equation}
\phi(t) = arg(z(t)) = tan^{-1}\left(\frac{\mathcal{H}[x(t)]}{x(t)}\right)
\end{equation}
representing the instantaneous phase of the signal. 

\begin{figure*}[h]
    \centering
    \includegraphics[width=0.9\linewidth]{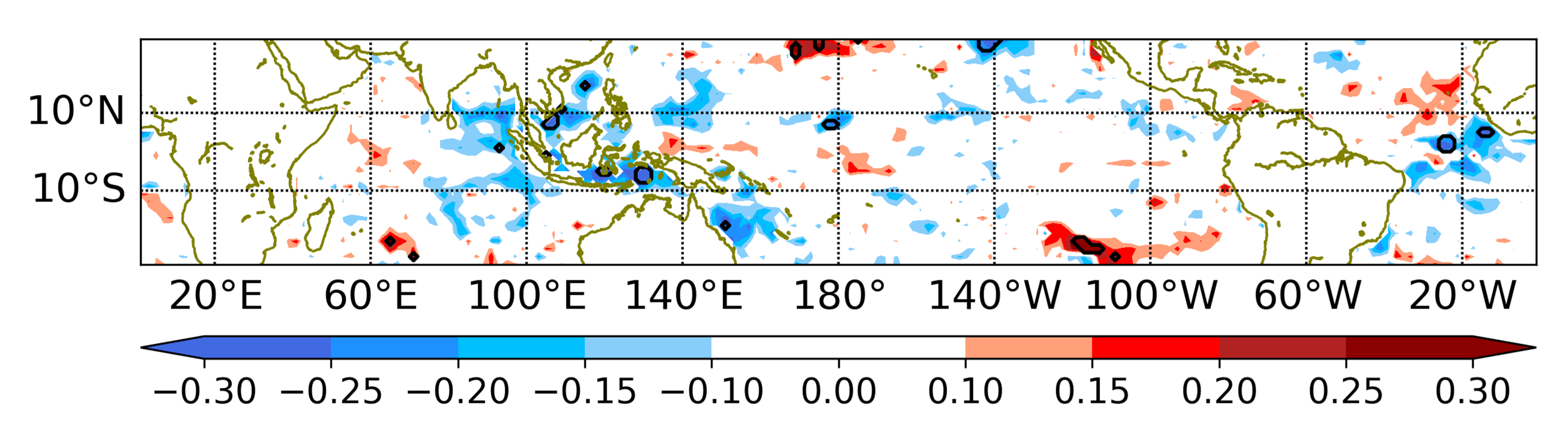}
    \caption{Pearson correlation between outgoing network strength (1941-2023) and OND rainfall anomalies (1942-2024). Black contours denote regions where the correlation is statistically significant at the 95\% confidence level based on the p-value statistic and phase-locking value (PLV).
}
    \label{supp_network_after_plv}
\end{figure*}

Therefore, for two oscillatory signals containing $N$ data points, the PLV is computed as
\begin{equation}
{PLV} = \left| \frac{1}{N} 
\sum_{t=1}^{N}
e^{i\,\Delta\phi(t)} \right|,
\end{equation}
where $\Delta\phi(t) = \phi_1(t) - \phi_2(t)$, and $\phi_1(t)$ and $\phi_2(t)$ are the instantaneous phases of $x_1(t)$ and $x_2(t)$, respectively. Pearson correlations filtered using PLV are presented in Fig. \ref{supp_network_after_plv}.

\section{Growth rate of Errors in ISMR} \label{Growth rate of Errors in ISMR}
\begin{figure*}[h]
    \centering
    \includegraphics[width=0.9\linewidth]{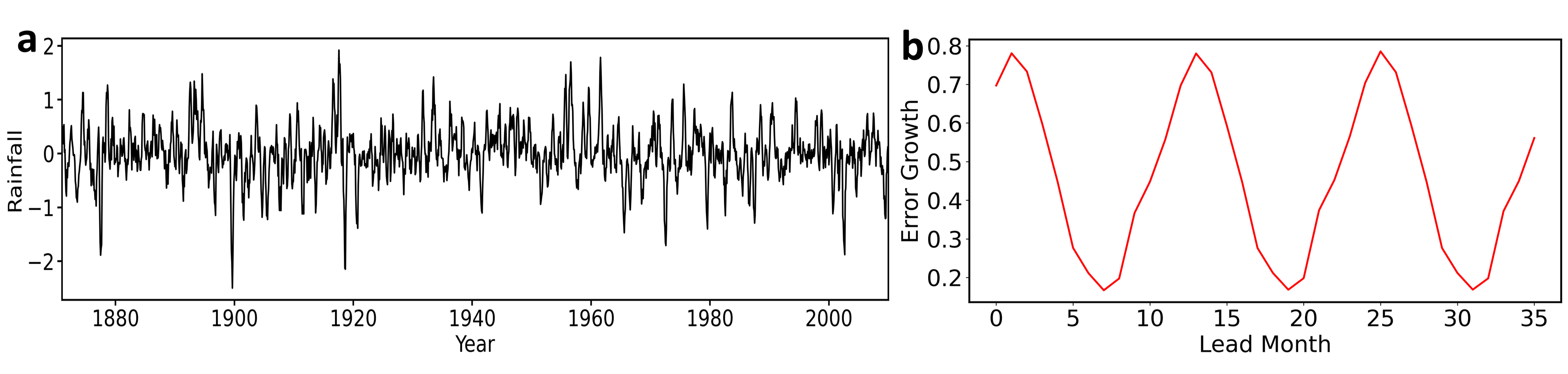}
    \caption{(a) 5-month running mean of monthly rainfall anomalies for the all-India region. (b) Growth rate errors in ISMR as a function of the lead forecast month.
}
    \label{error_growth}
\end{figure*}

Figure \ref{error_growth}a presents the 5-month running mean of monthly rainfall anomalies from 1871-2010, for the all-India region using the data set of \cite{parthasarathy1994all}. Figure \ref{error_growth}b presents the growth rate of errors in the all-India summer monsoon rainfall (ISMR) following the method of \cite{goswami2003potential}. Since the error growth here is dominated by an annual cycle, we present the growth of errors starting from the peak monsoon season only. From Fig. \ref{error_growth}b, we indicate that the deterministic limit of predictability of all India rainfall is about 6 months. 




\end{document}